\documentstyle[aps,preprint,epsf]{revtex}

\newcommand{\be}{\begin{eqnarray}} 
\newcommand{\ee}{\end{eqnarray}}       
\begin{document}

   \title{Semi-inclusive deep inelastic lepton scattering off complex nuclei} 
 \draft
   \author{C. Ciofi degli Atti,  
   L.P. Kaptari\thanks{On leave from Bogoliubov Laboratory  
   of Theoretical Physics, 
   JINR, 141980, Dubna, Moscow reg., Russia }} 
 
   \address{ 
   Department of Physics, University of Perugia and\\ 
   Istituto Nazionale di Fisica  Nucleare, Sezione di Perugia,\\ Via A. 
   Pascoli, 
   I-06100 Perugia, Italy}  
 
   \author{S. Scopetta} 
 
   \address{ Departament de Fisica 
   Te\`orica, Universitat de Val\`encia,\\ 46100 Burjassot, Val\`encia, Spain} 
    \date{\today}
   \maketitle 
 
   \begin{abstract}   
    
   It is shown that in  semi-inclusive deep inelastic scattering (DIS)
   of electrons off a complex nucleus $A$, 
   the detection, in coincidence with the scattered electron, 
   of a nucleus $(A-1)$ in the ground state, as well as of a nucleon 
   and a nucleus $(A-2)$,  also in the ground state, 
   may provide unique information on several long standing  
   problems, such as : 
$i)$  
the nature and the relevance of  the final state interaction in DIS; 
$ii)$ the validity of the spectator 
mechanism in DIS;  $iii)$ the   
 medium induced modifications of the nucleon structure function; $iv)$ the origin of the $EMC$ effect. 
   \end{abstract} 
 
%\leftline{KEYWORDS: semi--inclusive reactions; nucleon structure functions; 
%medium effects}  
 
%\newpage 
%\setcounter{page}{1} 
\section{Introduction} 
  
 In spite of many experimental and theoretical efforts  
(for a recent review see \cite{arneodo}),  
the origin of the nuclear EMC effect has not yet  been fully clarified,  
and the problem as to whether the quark distributions of 
nucleons undergo deformations due to 
the nuclear medium remains open. 
Understanding 
 the origin of the EMC effect  would be of great relevance in many 
respects; consider,  for example, that 
 most   QCD sum rules and predictions require  
  the knowledge of the neutron quark distributions, which 
 can only be extracted  from  nuclear experiments;  
 this implies, from one side, a reliable  knowledge of  various nuclear 
quantities, such as  the nucleon removal energy and momentum   
distributions, and,  from the other side,  a proper  treatment 
of the lepton-nucleus reaction mechanism, including 
the effect  
 of  final state interaction (FSI) of the leptoproduced 
hadrons with the nuclear medium. 
 Since the $Q^2$ and $x$-dependences of the EMC effect is smooth, 
 the measurements of the nuclear quark distributions 
 in inclusive deep inelastic scattering (DIS) processes  
have not yet established enough constraints to 
 distinguish between different theoretical approaches.  
In order to progress in this field, 
one should go beyond inclusive experiments, e.g.  by considering 
semi-inclusive experiments in which another particle  is detected in 
coincidence with the scattered electron. Most of theoretical 
studies in this field concentrated on the process $D(e,e'N)X$, where $D$ 
denotes the deuteron, $N$ a nucleon, and $X$ the undetected hadronic state. 
Current theoretical models of this process are based  
upon the impulse approximation (IA) (also called    
{\it the spectator model} ), according to which $X$ results 
from DIS on one of the two nucleons in the deuteron, 
with $N$ recoiling without interacting with $X$ 
and being  detected in coincidence  with 
the scattered electron (for an exhaustive review see 
\cite{fs}). The model has been improved by introducing 
FSI \cite{fsi}, as well as by considering deviations from 
the spectator model by assuming that the detected nucleon originates 
from quark hadronisation \cite{dieper,ciofisim}. The semi-inclusive process on the deuteron $D(e,e'N)X$, on which  experimental data 
will soon be available  \cite{semiexp}, could  not 
only clarify   the origin of the EMC effect, but,  as illustrated in   
\cite{simula},  
  could also provide more reliable information  
on the neutron structure function. 

The spectator model has also been extended 
to complex nuclei by 
considering the process $A(e,e'N)X$, and by assuming that DIS 
occurs on a nucleon 
of a correlated pair, with the second nucleon 
$N$ recoiling and being detected in 
coincidence with the scattered electron \cite{ciofisim}.  
In the present paper two  new types of semi-inclusive  
processes on complex nuclei will be considered,  
 namely: i) the process $A(e,e'(A-1))X$, in 
  which DIS occurs on a mean-field,  
low-momentum nucleon, and  
  the nucleus $(A-1)$ recoils   with low momentum and low excitation energy 
and is detected in coincidence with the scattered electron (note that for $A=2$ such a process   
 coincides with the 
process $D(e,e'N)X$ discussed previously);  
ii) the process $A(e,e'N{_2}(A-2))X$, in
which DIS occurs  
on a high momentum nucleon $N_1$ of a correlated pair, and the 
  nucleon $N_2$ and the   
nucleus $A-2$ recoil   with high and low momenta, respectively, and 
are detected in coincidence with the scattered electron.  
 It will be shown that these  processes  exhibit 
 a series of very interesting features which could in principle provide  useful insight on the following  basic issues: 
$i)$  
the nature and the relevance of  FSI in DIS; $ii)$ the validity of the spectator 
mechanism leading to the cross section (\ref{crosa-1});  $iii)$ the   
 medium induced modifications of the nucleon structure function; $iv)$ the origin of the $EMC$ effect. 
For the above reasons, the semi-inclusive processes we will consider are worth being theoretically analysed, even though their experimental investigation represents a difficult task.  It should 
be emphasised, in this respect, that the first version of the present  
paper\cite{nasharchiv} was motivated by the discussions 
on the feasibility  of an electron-ion collider,  
where the detection of various nuclear fragments  
resulting from DIS, could in principle be  possible \cite{hera,gsi}. 
 
Our paper is organised as follows: in Section II the basic  
nuclear quantities which enter the problem, viz. the one-body and  
two-body nuclear Spectral Functions 
are briefly discussed; the cross section for the process  
$A(e,e'(A-1))X$ is presented in Section III, where  
the possibilities offered by the process to experimentally  
check the validity of the spectator mechanism and the  
properties of the structure function of a mean-field,  
{\it weakly bound} nucleon, are discussed; the cross section for the process 
 $A(e,e'N{_2}(A-2))X$, and  
how this process can be used to investigate the spectator  
model and the properties of the structure function of a  
{\it deeply bound} nucleon, are discussed in Section IV; the  
{\it local} EMC effect, i.e. the separate contribution to  
the EMC effect of  nucleons  
having different binding in the nucleus, is discussed in  
Section V; the Summary and Conclusions are presented in  
Section VI. Appendix A contains the derivation of the  cross sections for  
both processes. 
   
\section{The nuclear spectral function} 
   
In order to make clear the nuclear  
physics aspects underlying the above processes,   few basic concepts 
about the  relationships between the  
 nucleon momentum distributions   in the 
   parent nucleus $A$ and the excitation energy of daughter nuclei  
   $(A-1)$ and   $(A-2)$ in semi-inclusive processes, will be recalled. 
    The nucleon Spectral Function $P_{N_1}(|\vec p_1|,E)$ represents the joint probability 
to have in the parent nucleus  a  
   nucleon with momentum  $|\vec p_1|$ and removal energy $E$
   \begin{eqnarray} 
   && 
   P_{N_1}(|\vec p_1|,E)= 
   \langle \Psi_A^0\,|\, a^+_{\vec p_1}\delta \left ( E- ( H_A-E_A^0)  \right ) 
   \, a_{\vec p_1}\,|\Psi_A^0\rangle =\nonumber\\[1mm] 
   && 
   \sum_f \left |  
   \langle \vec p_1, \Psi_{A-1}^f\,|\Psi^0_A 
   \rangle 
   \right |^2\delta \left (E-( E_{A-1}^f-E_A^0)\right), 
   \label{spectr1} 
   \end{eqnarray} 
   where $a^+_{\vec p_1}$ and $a_{\vec p_1}$ are creation and annhilitation operators, $H_A$ is the nuclear Hamiltonian, $E_A^0$ ($\Psi_A^0$)is the ground state energy (wave function) of $A$, and $E_{A-1}^f=E_{A-1}^0 + E_{A-1}^*$ ($\Psi_{A-1}^f$)is the intrinsic energy (wave function) of $A-1$, whose ground state energy is  $E_{A-1}^0$. Thus, the nucleon removal energy $ E = E_{A-1}^f - E_A^0 =
   M_{A-1} +M - M_A + E^*_{A-1}$ (where $M_i$ is the mass of system $i$) 
   is   
   the energy required to remove a nucleon 
   from $A$ leaving $(A-1)$ with excitation energy $E^*_{A-1}$. 
    
    A common   representation of the spectral function is as follows 
  (omitting unnecessary here indices and summations) \cite{ciofispec} 
   \be 
 P_{N_1}^A(|\vec p_1|,E) =  P_0(|\vec p_1|,E)+P_1(|\vec p_1|,E)
\label{spectrrap}
\ee
where
\be
P_0(|\vec p_1|,E)\label{eq1} =
   \sum\limits_{\alpha < F}n^A_\alpha(|\vec p_1|)\delta(E-\varepsilon_\alpha)
\label{P0}
\ee
and
\be
P_1(|\vec p_1|,E)  =   
   {1 \over (2 \pi)^3} {1 \over 2 J_0 + 1} \sum_{M_0 \sigma} 
       ~ \sum_{f \neq \alpha} \left | \int d \vec{r} ~ e^{i\vec{p_1}  
   \cdot \vec{r}} ~ 
       G_{f0} 
   %^{\sigma} 
   (\vec{r}) \right |^2 ~ \delta[E - (E_{A-1}^f - E_A)]
\label{P1} 
   \ee 
   In the above equations  
   $F$ denotes the Fermi level, $n^A_\alpha(|\vec p_1|)$ is the momentum 
   distribution 
   of a bound shell model state with eigenvalue 
   $\varepsilon_\alpha > 0$, and $G_{f0}$ is the overlap between the 
   wave functions of the ground state of the parent $A$ and the state $f$ 
   of the daughter $(A-1)$ (see for details ref.~\cite{ciofi-fs,ciofi1,mar}).  
   The quantity 
   $P_0(|\vec p_1|,E)$, represents the shell model contribution to the Spectral Function, where the occupation numbers of the shell model states below the Fermi sea are given by $N_{\alpha}= \int {{d\vec p_1}n^A_\alpha(|\vec p_1|)} < 1$, whereas $P_1(|\vec p_1|,E)$ provides the contribution from correlations, which deplete the shell model states $\alpha < F$. 
   The so called Momentum Sum Rule links the spectral function to the  
   nucleon momentum distribution, viz. 
   \be 
   n^A(|\vec p_1|)=\int\limits_{E_{\rm min}}^\infty\,P_{N_1}^A(|\vec p_1|,E)    dE = 
   \sum\limits_{\alpha < F}n^A_\alpha(|\vec p_1|) 
   +   
   %{1 \over 2 \pi^2} {1 \over 2 J_0 + 1} \sum_{M_0 \sigma}  
   ~ \sum_{f \neq \alpha} 
   %    \{\alpha < \alpha_F \}}  
   \left | \int d \vec{r} ~ e^{i\vec{p_1} \cdot 
       \vec{r}} ~ G_{f0} 
   %^{\sigma} 
   (\vec{r}) \right |^2~, 
   \label{eq2} 
   \ee 
where $E_{min}=E_{A-1}-E_A$. 
   It can therefore be seen that  
$ 
n^A_0(|\vec p_1|) \equiv 
\displaystyle\sum\limits_{\alpha 
   < F} 
   n^A_\alpha(|\vec p_1|)=  
\int\limits_{E_{\rm min}}^\infty\,P^A_0(|\vec p_1|,E) dE , 
$  
represents the momentum distribution in the parent, when 
the  
   daughter is either in the ground state or in hole states 
   of the parent, 
   whereas  
$ 
n^A_1(
|\vec p_1|) \equiv n^A(
|\vec p_1|)-n^A_0(
|\vec p_1|)= 
\int\limits_{E_{\rm min}}^\infty\,P^A_1(|\vec p_1|,E) dE 
$  
   represents the momentum 
   distribution 
   in the parent, when the daughter is left in highly excited states,  
   with at least one particle in the continuum; this 
   means 
  that $n^A_0(
|\vec p_1|)$ is the momentum distribution of weakly bound (shell-model) 
     nucleons, while $n^A_1(
|\vec p_1|)$ is the momentum distributions of deeply bound nucleons 
   generated by N-N correlations. 
   A realistic model for the latter leads to 
   the following form of 
   the corresponding spectral function $P^A_1(
|\vec p_1|,E)$ 
   \cite{ciofi-fs,ciofi1,mar} 
   \be  
   && 
   P^A_{1}(
|\vec p_1|,E)= \label{sfcorr}\\[3mm] 
   && 
   \int d^3k_{cm} n^A_{rel}\left (|\vec p_1 -\vec p_{cm}/2|\right ) 
    n^A_{cm} (|\vec p_{cm}|) 
   \delta \left [ E -E_{thr}^{(2)} - 
   \frac{(A-2)}{2M(A-1)}\cdot \left ( \vec p_1 - \frac{(A-1)\vec 
   p_{cm}}{(A-2)} 
   \right )^2\right ], 
   \nonumber 
   \ee 
   where $n^A_{rel}$ and $n^A_{cm}$ are, respectively,   
   the relative and Center of Mass 
   momentum distributions of a correlated pair.  
 
 %The representation of the  
 %function $P^A(p_1,E)$ and  
%of the momentum distribution $n^A(
%|\vec p_1|)$ in form of convolution 
%integrals, eqs. (\ref{eq1}) - (\ref{sfcorr}), means the assumption that  
%the high 
%momentum and high removal energy parts of the spectral function are 
%solely governed by those ground state configurations where two nucleons 
%are very close and form a correlated pair, the pair itself being 
% relatively far from the remaining $A-2$ nucleons. Consequently, 
%the relative momentum of the correlated nucleons is large 
%($p_{rel} \ge 0.3 GeV/c$ ) whereas the CM momentum of the pair is low. 
It has been shown \cite{ciofi1} that such a model 
 satisfactorily reproduces the nuclear spectral functions calculated 
within many-body approaches with realistic $NN$ interaction and describes 
fairly well the quasi elastic inclusive $A(e,e')X$ processes.  

Within the spectator model, the process $A(e,e'(A-1))X$  is directly proportional to the one-nucleon spectral function, whereas
 the 
process   
$A(e,e'N(A-2))X$ is proportional to the two-nucleon spectral function, which is defined as follows 
\begin{eqnarray} 
   && 
   P_{N_1N_2}(\vec p_2,\vec p_1,E^{(2)})= 
   \langle \Psi_A^0\,|\, a^+_{\vec p_1}a^+_{\vec p_2} 
   \delta \left ( E^{(2)} -  ( H_{A-2} - E_A)  \right ) 
   \, a_{\vec p_2}a_{\vec p_1}\,|\Psi_A^0\rangle =\nonumber\\[1mm] 
   && 
   \sum_f \left |  
   \langle \vec p_1,\vec p_2, \Psi_{A-2}^f\,|\Psi^0_A 
   \rangle 
   \right |^2 
   \delta \left (E^{(2)} - ( E_{A-2}^f - E_A)\right), 
   \label{spectr2} 
   \end{eqnarray} 
where $E^{(2)}= E^{(2)}_{th} + E_{A-2}^*$ is the two-nucleon 
removal energy, $  E_{A-2}^*$ is the intrinsic 
 excitation energy of the $A-2$ system, and  
$E^{(2)}_{th} = 2M + M_{A-2} -M_A$ the two- 
nucleon break-up threshold. If one adheres to the model   
 leading to Eq. (\ref{sfcorr}), 
 the correlated part of the two-nucleon spectral function  
 can be written as follows\cite{ciofisim}: 
 \begin{eqnarray}&& 
 P_{N_1N_2}(\vec p_1,\vec p_2,E^{(2)})= 
  \,n^A_{cm}(|\vec P_{A-2}|)\, n^A_{rel.}(|\vec p_2 +\vec P_{A-2}/2 |) 
  \delta \left ( E^{(2)}-  E^{(2)}_{th}\right ) 
  \label{ster} 
 \end{eqnarray} 
\section{The {\bf \it A\lowercase{(e,e'}(A-1))X} process} 
 
In Impulse Approximation, the process $A(e,e'(A-1))X$ (depicted  
 Fig.\ref{fig1}a), represents the absorption of the the virtual photon  by a quark of a shell-model nucleon, followed by the recoil of the nucleus  $A-1$ 
 in a low momentum ,${\vec P_{A-1}}$, and low excitation energy, $E_{A-1}^*$,  state ($E_{A-1}^* \simeq 0$ or $\simeq$  shell-model   hole state energy of the target);the scattered electron and the nucleus $(A-1)$ are detected in coincidence . The aim for studying such a process 
  is twofold:
\begin{itemize}
\item [ i)] to investigate the  
 nature of the final state interaction (FSI) of the hit  
 quark with the surrounding nuclear 
 medium; as a matter of fact,  the observation of a nucleus 
 $(A-1)$ in the ground state (or in a low shell model excited states) 
 would represent obvious  evidence that the leptoproduced hadrons propagated
through the nucleus $(A-1)$ without strong FSI. Therefore, the number of observed $(A-1)$ systems and its variation with $A$ could provide   
 important information on e.g. the hadronization length in  
 the medium ;
\item  [ii)] to investigate 
 the $A$-dependence of possible medium induced modifications of the 
 DIS structure function of weakly bound nucleons. 
 \end{itemize} 
In IA   
the differential cross section in the laboratory  system  
has the following form  (see Appendix A)~\cite{nasharchiv} 
   \begin{eqnarray} 
   &&\!\!\!\!\!\! \sigma^A_1 (x_{Bj},Q^2,\vec P_{A-1})\equiv\sigma^A_1= 
  \frac{d\sigma^A}{d x_{Bj} d Q^2  d \vec   P_{A-1}}\nonumber\\&& 
   =  
   K^A( x_{Bj},Q^2,y_A,z_1^{(A)}) z_1^{(A)}  
   F_2^{N/A}(x_A,Q^2,p_1^2)\,n^A_0(|\vec P_{A-1}|), 
   \label{crosa-1} 
   \end{eqnarray} 
   where: $Q^2 =-q^2= -(k_e-k_e')^2 = \vec q^{\,\,2} - \nu^2=4 {\cal E}_e  
{\cal E}_e' sin^2 {\theta \over 2}$ is the 4-momentum transfer  
(with $\vec q = \vec k_e - \vec k_{e'}$, $\nu= {\cal E}_e - 
{\cal E}_e' $ and $ \theta \equiv \theta_{\widehat{\vec k_e  \vec k_{e'}}}$); 
$  x_{Bj} = Q^2/2M\nu $ is the Bjorken scaling variable;  
 $p_1 \equiv(p_{10},\vec {p_1})$, 
 with  $\vec {p_1} \equiv - \vec P_{A-1}
$, is the four momentum of the 
 nucleon; 
$F_2^{N/A}$ is the DIS structure function of the nucleon 
$N$ in the nucleus $A$; 
$n_0^A(|\vec P_{A-1}|)$ is the 3-momentum distribution 
of the bound nucleon; 
  $K^A( x_{Bj},Q^2,y_A, z_1^{(A)}) $ is the following kinematical factor 
   \begin{eqnarray} 
   && 
   K^A( x_{Bj},Q^2,y_A,z_1^{(A)})= 
   \frac{4\alpha^2}{Q^4} \frac {\pi}{x_{Bj}}\cdot  
     \left( \frac{y}{y_A}\right)^2  
   \left[\frac{y_{A}^2}{2} + (1-y_A) -  
   \frac{p_1^2x_{Bj}^2 y_A^2}{z_1^{(A)2}Q^2}\right ]~, 
   \label{ka} 
   \end{eqnarray} 
and 
\begin{eqnarray} 
&& 
y=\nu/{\cal E}_e ~, \,\,\,\,  y_A = (p_1\cdot q)/(p_1\cdot k_e) 
\label{ydef}\\ 
&& 
 x_A = {x_{Bj} \over z_1^{(A)}}, \quad 
z_1^{(A)} = {p_1 \cdot q \over M \nu}~. 
\label{adef} 
\end{eqnarray} 
 Nuclear effects in  Eq. (\ref{crosa-1}) are generated    
 by the nucleon momentum distribution $n_0^A(|\vec P_{A-1}|)$,  
  and by the quantities $y_A$ and $z_1^{(A)}$, which  
  differ from the corresponding quantities for a free 
  nucleon  ($y=\nu/{\cal E}_e$ and $z_1^{(N)}=1$),  if the off mass shellness of 
  the nucleon  ($p_1^2\neq M^2$ ) generated  by nuclear binding is 
  taken into account.  
   Equation (\ref{crosa-1})  
    is valid for  finite values of 
   $Q^2$, and for $A=2$  agrees with the expression  
     used  in refs. \cite{simula,fs}  
   (note, that in ref. \cite{simula} the quantity 
    $D^N=K^A/K^N$ has been 
   used, $K^N$ being the quantity (\ref{ka})   
    for a free nucleon, which will be discussed later on).

In this paper, we follow the usual procedure consisting of disregarding the explicit dependence of  
$F_2^{N/A}$ upon $p_1^2$, and choose the form of $F_2^{N/A}$ 
to be the same as for the free nucleon; within such an approach, the effect of the nuclear  
medium will be  considered within two main models: 
 
$i)$ {\underline{\sl the x-rescaling model}}, which directly follows from the 
convolution formula of inclusive scattering, leading to   energy 
conservation at the hadronic vertex in Fig.1, i.e.
 
\begin{equation} 
 p_{10}= M_A - \sqrt{ (M_{A-1} + E_{A-1}^*)^2 + \vec P_{A-1}^2}~, 
\label{p10} 
\end{equation} 
which, when placed in eqs. (\ref{ydef}) and (\ref{adef}),  
leads to the following  structure function for a bound nucleon
 
\begin{equation} 
F_2^{N/A}(x_A,Q^2,p_1^2)=F_2^{N/A}\left (\frac{x_{Bj}}{z_1^{(A)}},Q^2 \right) 
\label{f2na} 
\end{equation} 
 
with 
 
\begin{equation} 
z_1^{(A)} = 
(p_{10}+|\vec P_{A-1} |\eta \cos\theta_{\widehat{\vec P_{A-1} \vec q} }) /M~, 
\label{zan} 
\end{equation} 
 
and 
 
\begin{equation} 
\eta = |\vec q|/\nu\,=\sqrt{1+\frac{4M^2x_{Bj}^2}{Q^2}}~. 
\label{eta} 
\end{equation} 
 
Since the $(A-1)$ system is detected in a low excited
state ($E_{A-1}^* \simeq 0$) and with low momentum  
($|\vec P_{A-1}|<<M_{A-1}$), Eq. (\ref{p10}) can be 
safely replaced by 
 
\begin{equation} 
p_{10} \simeq (M-E_{min})- {|\vec P_{A-1}|^2 \over 2M_{A-1}}~, 
\label{p10a} 
\end{equation}
Eq. (\ref{zan}) then becomes 
 
\begin{equation} 
z_1^{(A)} \simeq 1 - {E_{min} \over M} - {|\vec P_{A-1}|^2 \over 2 M M_{A-1}} 
+ \frac{\eta}{M}  |\vec P_{A-1}| cos \theta_{\widehat{\vec P_{A-1} \vec q}}~ 
\label{zac} 
\end{equation} 
and for  a heavy nucleus, for which the recoil term in (\ref{zac}) 
is negligibly small, one has 
\begin{equation} 
z_1^{(A)} \simeq 1 - {E_{min} \over M} + 
 \frac{\eta}{M}  |\vec P_{A-1}| cos \theta_{\widehat{\vec P_{A-1} \vec q}}~. 
\label{zaf} 
\end{equation} 
 Moreover, being ${E_{min} \over M} << 1$, it can be concluded that the structure functions 
(\ref{f2na}) will exhibit almost no $A$-dependent effects, apart from 
the case of the   
few nucleon systems ($A$=2,3,4), for which the recoil term in (\ref{zac}) 
cannot be disregarded. 
In the Bjorken limit 
 ( $Q^2\to\infty,\,\, \nu\to\infty$, 
$x_{Bj}={\rm const,\,\, \nu\sim\ |\vec q| }$), $\eta\to 1$)
\begin{equation}
z_1^{(A)}= (p_{10}+|\vec P_{A-1}| \cos \theta_{\widehat{\vec P_{A-1} 
\vec q}})/M
\label{zab}.
\end{equation} 
 Note that Eq.  
(\ref{zab}) can also be written as  ( 
$E_{A-1}^* = 0$ in the processes we are considering)  
\begin{equation} 
z_1^{(A)}=\frac{M_A}{M} -\frac{M_{A-1}z_{A-1}}{M} 
\label{z1} 
\end{equation} 
where 
\begin{equation} 
z_{A-1}=  \frac{ 
\sqrt{\vec P_{A-1}^2 +M_{A-1}^2} - |\vec P_{A-1}| 
\cos\theta_{\widehat{\vec P_{A-1} \vec q} } }{M_{A-1}} 
 \end{equation} 
 is the light cone momentum of the $A-1$ recoiling 
 nucleus. 
  Eq. (\ref{z1}) is nothing but the energy conservation of the process  
 \begin{equation} 
\nu +M_A=\sqrt{M_X^2+(\vec p_1+\vec q)^2}+\sqrt{M_{A-1}^2+{\vec p_1}^2} 
 \label{conserv} 
 \end{equation}  
 in the Bjorken limit, where $M_X$ is the invariant mass of the 
 produced hadronic state $X$; 
 in the case of the deuteron, the term  
 $\displaystyle\frac{|E_{min}|}{M}$ can be disregarded, so that 
$M_A/M \simeq 2$  and  the well known relation 
 $z_1^{(2)}=2-z_2$, where  
 $z_2 = (\sqrt{|\vec p_2|^2 +M^2}-|\vec p_2|\cos  
\theta_{\widehat{\vec p_2 \vec q}})/M$, 
and $\vec p_2$ is the momentum of the recoiling nucleon, is recovered.
 
$ii)$ {\underline{\sl the $Q^2-$rescaling model}} \cite{close}, which is based 
on the idea of a medium modification of the $Q^2-$evolution equations 
of $QCD$, leading to

\begin{equation} 
F_2^{N/A}(x,Q^2)=F_2^{N}(x, \xi_A (Q^2)Q^2)~, 
\label{f2nar} 
\end{equation} 
where the $Q^2$ dependence of the quantity $\xi_A (Q^2)$ is determined 
so as to satisfy the $QCD$ evolution equations on both sides of   
(\ref{f2nar}), with the additional hypothesis that the quark 
confinement radius for a bound nucleon ($\lambda_A$) is larger than 
that for a free nucleon ($\lambda_N$), according to the ansatz 
 
\begin{equation} 
{\lambda_A^2 \over \lambda_N^2} = {\mu_N^2 \over \mu_A^2} = \xi_A (\mu_A^2)~, 
\label{lamb} 
\end{equation} 
where $\mu_A$ and $\mu_N$ are the lower momentum cutoffs for 
the bound and free nucleons, respectively. The following relation can then  
be obtained 
 
\begin{equation} 
\xi_A(Q^2)= \left (\lambda_A^2 \over \lambda_N^2 \right ) 
^{ ln( Q^2/ \Lambda_{QCD}^2) \over ln( \mu_A^2 /\Lambda_{QCD}^2) }~, 
\label{csia} 
\end{equation} 
where $\Lambda_{QCD}$ is the universal $QCD$ scale parameter.

To sum up, in the $Q^2$-rescaling model an explicit $A$ dependence is 
provided by Eq. (\ref{csia}) whereas, in 
the $x$-rescaling model, the $A$-dependence of $F_2^{N/A}$ 
is generated implicitly   
by the momentum $|\vec P_{A-1}|$ of the detected $A-1$ system 
(cf. Eq. (\ref{zac})).

We will now discuss  a series of processes, 
which could in principle provide useful insight on the following  basic issues: 
$i)$  
the nature and the relevance of  FSI in DIS; $ii)$ the validity of the spectator 
mechanism leading to the cross section (\ref{crosa-1});  $iii)$ the   
 medium induced modifications of the nucleon structure function; $iv)$ the origin of the $EMC$ effect.

\subsection{Checking the spectator mechanism in the semi-inclusive process 
$A(e,e'(A-1))X$} 
 
The validity of the 
spectator mechanism could  experimentally be checked in the following way. 
Let us consider the cross section (\ref{crosa-1}) for two 
different nuclei $A$ and $A'$, and the same values of 
$x_{Bj}$, $Q^2$ and $|\vec P_{A-1}|=|\vec P_{A'-1}|$. Consider now the ratio 
 
   \begin{eqnarray} 
   && 
   R(x_{Bj},Q^2,|\vec P_{A-1}|,z_1^{(A)},z_1^{(A')},y_A ,y_{A'})= 
  \frac{\sigma^A_1 ( x_{Bj},Q^2,|\vec P_{A-1}|,z_1^{(A)},y_A ) } 
             {\sigma^{A'}_1 ( x_{Bj},Q^2,|\vec P_{A-1}|,z_1^{(A')},y_{A'} ) }= 
   \nonumber \\ 
   & = &   \frac{K^A}{K_{A'}} 
   \frac{z_1^{(A)} F_2^{N/A}(x_A,Q^2,p_1^2)}{z_1^{(A')} 
   F_2^{N/A'}(x_{A'},Q^2,p_1^2) } 
   \frac{n_0^A(|\vec P_{A-1}|)}{n_0^{A'}(|\vec P_{A-1}|)}~, 
   \label{ratioa-1} 
   \end{eqnarray} 
with $y_A$ and $z_1^{(A)}$ defined in Eqs. (\ref{ydef}) and (\ref{adef}), 
 respectively. For reasons that would be clear later on,  
 our aim is to get rid as much as possible 
of the various $A$ and $A'$ dependencies appearing in (\ref{ratioa-1}) , except the ones provided by the 
nucleon  
momentum distributions.  
The dependence    upon $A$ and $A'$  
is contained in the quantities $z_1^{(A)}$, $x_A$, $K^A$, and 
$n_0^A(|{\vec P_{A-1}}|)$; 
  in order to get rid of the $A$-dependence due to the first 
  three quantities 
  let us   consider coplanar kinematics, i.e. 
\be 
&& 
y_A=y\cdot\frac{ p_{10}+\eta |\vec P_{A-1}|\cos  
\theta_{  \widehat{ \vec P_{A-1} \vec q} }  } 
{ p_{10}+\eta |\vec P_{A-1}|\cos  
\theta_{ \widehat{ \vec P_{A-1} \vec k_e} } }~, 
\quad  
\label{add1} 
\ee 
with 
\be 
&& 
\cos\theta_{ \widehat{ \vec P_{A-1} \vec q}} = - \cos(  
\theta_{\widehat {\vec P_{A-1} \vec k_e}} 
+\theta_{ \widehat {\vec q \vec k_e}});\quad 
\cos\theta_{ \widehat {\vec q \vec k_e} } =   
\left (1+\frac{Mx_{Bj}} {{\cal E}_k}\right )/\eta~. 
\label{add2} 
\ee 
In the Bjorken limit
$\eta\to 1$, $z_1^{(A)}= (p_{10}+|\vec P_{A-1}| \cos \theta_{\widehat{\vec P_{A-1} 
\vec q}})/M$ 
   (cf. Eq. (\ref{zan})),  
$\theta_{ \widehat{ \vec P_{A-1} \vec q}}\to   
\theta_{ \widehat{\vec P_{A-1} \vec k_e}}$,   
$y_A\to y$, and 
   \begin{eqnarray} 
   && 
   K^A( x_{Bj},Q^2,y_A,z_1^{(A)}) \rightarrow  K^N( x_{Bj},Q^2,y) = 
   \frac{4\alpha^2}{Q^4} \frac {\pi}{x_{Bj}}\cdot  
   \left[\frac{y^2}{2} + 1 - y - { Q^2 \over 4 {\cal E}_e^2}   
   \right ]~, 
   \label{kap} 
   \end{eqnarray} 
where $K^N$ is nothing but the trivial kinematic factor appearing  
in the DIS $eN$- inclusive cross section; 
the cross section (\ref{crosa-1}) thus  becomes  
 \begin{eqnarray} 
   &&\!\!\!\!\!\! 
\left (  \frac{d\sigma^A}{d x_{Bj} d Q^2  d \vec   P_{A-1}}\right )_{Bj} 
   =  
   K^N( x_{Bj},Q^2,y) z_1^{(A)} F_2^{N/A}(x_{Bj}/z_1^{(A)}, 
   Q^2)\,n^A_0(|\vec p_{A-1}|), 
   \label{crossBj} 
   \end{eqnarray} 
   with the $A$-dependence now appearing only in  $ F_2^{N/A}$,  
 $n^A_0(|\vec p_{A-1}|)$ and $z_1^{(A)}$. 
   The latter  dependence, however, can be 
   eliminated by considering that 
Eq. (\ref{zac}) reduces (due to $E_{min}/M <<1$ ) 
to  
$z_1^{(A)} \simeq 1 - {|\vec P_{A-1}|^2 \over 2 M M_{A-1}} 
+ \frac{ |\vec P_{A-1} | } {M} cos \theta_{\widehat{ \vec P_{A-1} \vec q}}~,$ 
so that by   by fixing $|\vec P_{A-1}|$, 
and properly changing 
$\theta_{ \widehat{\vec P_{A-1} \vec q} }$, the condition 
$z_1^{(A)} \sim z_1^{(A')}$ can easily be achieved. 
As a result, the Bjorken limit of Eq.(\ref{ratioa-1}) becomes 
\begin{eqnarray} 
 R_{Bj}(x_{Bj}/z_1^{(A)},Q^2,|\vec P_{A-1}|,A,A') & = & 
 \frac {F_2^{N/A}(x_{Bj}/z_1^{(A)},Q^2)} 
         {F_2^{N/A'}(x_{Bj}/z_1^{(A')},Q^2)} 
\frac{n_0^A(|\vec P_{A-1}|)}{n_0^{A'}(|\vec P_{A-1}|)} 
\rightarrow 
\\ \nonumber
& \rightarrow &
\frac{n_0^A(|\vec P_{A-1}|)}{n_0^{A'}(|\vec P_{A-1}|)}\equiv 
R(| {\vec P_{A-1}}\ |), 
  \label{eq5} 
\end{eqnarray} 
where the last step is strictly  valid  only within the 
$x$-rescaling model, for in the $Q^2$-rescaling model 
the additional $A$ and  $A'$-dependences appearing 
in  
$F_2^{N/A}(x, Q^2) =F_2^N(x, \xi_A(Q^2)Q^2)$ 
does not cancel out, being different in the numerator and the denominator; 
  such a dependence, however, is 
overwhelmed by the $A$-dependence of $n_0(\left | {\vec P_{A-1}}\right |)$ 
as it will be shown later on. 
  
We have thus obtained that in the 
Bjorken limit  the $A$ dependence 
of the ratio $R$ is entirely governed by the $A$ dependence  
of the nucleon momentum distribution $n_0^A(|\vec P_{A-1}|)$.  
 Since the latter exhibits a strong $A$ dependence for low values of 
$|\vec P_{A-1}|$, a plot of $R$ versus $|\vec P_{A-1}|$ should  
reproduce the   behaviour of $n_0^A(|\vec P_{A-1}|)$ which is 
fairly well known, so that   
the experimental observation of such a behaviour would 
represent a stringent test of the spectator mechanism  
independently of the model for $F_2^{N/A}$. 
  
Fig. 2 illustrates the expected behaviour of the ratio   
(\ref{eq5}) for $A=2$ and different values of $A'$.  The measurement 
of the quantity $R$ shown in Fig. 2  would imply the detection, 
in coincidence with scattered electrons, of backward recoiling , 
with  momentum $\vec P_{A-1}$,  protons, 
deuterons, $^3He$ and $^{12}C$  nuclei resulting from the processes 
$D(e,e'p)X$, $^3H(e,e'D)X$, $^4He(e,e'^3He)X$ and 
$^{12}C(e,e'^{11} C)X$, respectively, with  the DIS  
scattering processes supposed to occur on a neutron. \footnote { 
Note that the condition $z_1^{(A)}/z_1^{(A')} =1$ cannot be 
achieved if both  $\theta_{\widehat{\vec P_{A-1}  \vec q}}$ are  
fixed, so that in  Fig.2 $z_1^{(A)}/z_1^{(A')}$ 
is a function of $P_{A-1}$; however the     
   $P_{A-1}$-dependence of the quantity  
   $z_1^{(A)}F_2^{N/A}(x_{Bj}/z_1^{(A)})/ 
   z_1^{(A')}F_2^{N/A'}(x_{Bj}/z_1^{(A')})$ 
   is at most of the order 5 percents and the dependence 
   of $R(|\vec P_{A-1}|)$ upon $|\vec P_{A-1}|$ is  
   entirely provided by the momentum distributions.} 
   Since the results presented in Fig.2 were obtained in the 
   Bjorken limit, where $K^A=K^{A'}=K^N$, let us analyse  
   at which  value of $Q^2$  such an equality is fulfilled. 
   To this end, in  
   Fig. 3 the ratio $K^A/K^N$ is shown vs. $Q^2$ for  
    of $A=4$.   It can be seen that at $Q^2 \simeq 5 GeV^2$
 $K^A$  and $K^N$ differ by $5\%$ only. The cross sections corresponding
to the processes considered in Fig.2 are presented in Fig.4.

    From the results we have exhibited, it is clear that the 
    observation of recoiling nuclei in the ground state, 
    with a $|\vec P_{A-1}|$-dependence  similar to the one predicted  
    by the momentum distributions, would represent 
    a stringent check of the spectator mechanism, which, in turns, 
    would indicate the absence of significant FSI between the 
    lepto-produced hadronic states and the nuclear medium. 
   The experimental observation 
    of $(A-1)$ nuclei in the ground states would represent  
    a strong indication that the hadronization length 
 is larger    than the effective nuclear dimension, since if 
 the hit quark hadronizes inside the nucleus, the latter
    is expected to be strongly excited.
 Of particular relevance, in this respect,  would be  
 the processes  $^3He (^3H) (e,e'D)X$, for 
if FSI plays an important role,  the weakly bound  
final state  deuteron   will easily break down. 
It is clear, therefore, that 
the experimental observation of  
the exclusive   process $A(e,e'(A-1)_{gr})X$ 
is strong evidence of the smallness of FSI. 
Although recent calculations \cite{fsia} and experimental 
data \cite{adams} seem to indicate that FSI on  
a complex nucleus are small in semi inclusive DIS, 
particularly when the low momentum hadrons are detected 
backward, the situation is not  
clearly settled, and therefore the observation of protons 
and deuterons emitted backward in the processes 
$D(e,e'p)X$, $^3He(e,e'D)X$ 
with a $|\vec P_{A-1}|$ dependence shown in Fig.2, 
would represent strong indication of the absence of  FSI. 
 The situation here is different from the usually  
 investigated  semi-inclusive DIS processes $A(e,e'N)X$  
 where the detected nucleon 
 can originate not only from a correlated 
 pair, as originally suggested\cite{fs}, but from  competitive processes 
as well, 
 such as nucleon current and target fragmentations  \cite{dieper,ciofisim}. 

Let us now discuss the possibility to obtain   
information on the nucleon structure function of a weakly bound nucleon 
by means of the process $A(e,e'(A-1))X$.

\subsection{Investigating the structure functions of weakly bound 
nucleons by the process $A(e,e'(A-1))X$} 
 
Consider  the following quantity 
\begin{equation} 
  R^A(x_{Bj},x_{Bj}',Q^2,|\vec P_{A-1}|) \equiv  
  \frac{ \sigma_1^A (x_{Bj}, Q^2,|\vec P_{A-1}|,z_1^{(A)},y_A)}  
       { \sigma_1^A (x_{Bj}',Q^2,|\vec P_{A-1}|,z_1^{(A)},y_A)} 
  \label{x1x2} 
\end{equation}     
which represents the ratio between the cross section (\ref{crosa-1}) 
on the  nucleus $A$ considered at two different values of the Bjorken 
scaling variable. 
It is clear that all terms of (\ref{crosa-1}), 
but the nucleon structure functions, cancel out 
in the ratio,  and one has 
 
\begin{equation} 
  R^A(x_{Bj},x_{Bj}',z_1^{(A)},Q^2) = 
  { x_{Bj}' \over x_{Bj} }   
  \frac{ F_2^{N/A} \left ( { x_{Bj} \over z_1^{(A)}   }   , Q^2 \right ) }  
       { F_2^{N/A} \left ( { x_{Bj}' \over z_1^{(A)}}   , Q^2 \right ) } 
  \label{x1x2x} 
\end{equation} 
in the $x-$rescaling approach, and 
 
\begin{equation} 
  R^A( x_{Bj}, x_{Bj}', Q^2 ) = 
  { x_{Bj}' \over x_{Bj} }   
  { F_2^{N/A } \left ( x_{Bj},  \xi_A (Q^2) Q^2 \right )  
  \over  
    F_2^{N/A} \left ( x_{Bj}', \xi_A (Q^2) Q^2 \right ) }  
       = constant~, 
  \label{x1x2q} 
\end{equation} 
 in the $Q^2-$rescaling model. 
 Eqs. (\ref{x1x2x}) and (\ref{x1x2q}) will in general  
exhibit 
a different $|\vec P_{A-1}|$ dependence: Eq. (\ref{x1x2q}) will be  
a $|\vec P_{A-1}|$-independent constant  different for different 
nuclei, whereas Eq. (\ref{x1x2x}) will depend both upon  $A$ and  
$|\vec P_{A-1}|$, due to the dependence of $z_1^{(A)}$ upon 
$|\vec P_{A-1}|$ (cf. Eq. (\ref{zan}) with $\eta=1$).  
Let us consider the ratio (\ref{x1x2}) for the deuteron and for 
a complex nucleus; 
placing (\ref{p10a}) in (\ref{zan}), one obtains  $z_1^{(2)} 
\simeq 1 - {{\cal E}_D \over M} -  \frac{ |\vec P_{A-1}|^2}{2M^2}  
 +\frac{|\vec P_{A-1}|}{M} cos \theta_{ \widehat {\vec P_{A-1} \vec q} }$~ and a strong $|\vec P_{A-1}|$ dependence will originate from the recoil 
and the angle-dependent terms; for a complex nucleus, on gets 
$z_1^{(A)} \simeq 1 - { E_{min} \over M}  
 +\frac{|\vec P_{A-1}|}{M} cos \theta_{\widehat{\vec P_{A-1}\vec q}}$, 
  which  appreciably differs  from  
 unity  only for  $\theta_{\widehat{\vec P_{A-1} \vec q} }=180^o$ 
and/or large values of $|\vec P_{A-1}|$. 
Thus, the $|\vec P_{A-1}|$-dependence of the ratio (\ref{x1x2x}) 
can be changed by 
varying the dependence of $z_1^{(A)}$ upon $|\vec P_{A-1}|$; in such 
a way,  ${x_{Bj} \over z_1^{(A)}} \neq  
{x_{Bj}'\over z_1^{(A)}}$   
and  $R^A$ will differ from a constant. The ratio (\ref{x1x2x}), 
for $A=2$ and $A=40$, is shown in Figs.\ref{fig5} and \ref{fig6} 
in correspondence of two values of the emission angle  
$\theta_{\widehat{\vec P_{A-1} \vec q} }$ 
of the nucleus $A-1$ ( 
$\theta_{\widehat{\vec P_{A-1} \vec q} } 
=90^o$ and $180^o$), 
and $x_{Bj}=0.2$ and  $x_{Bj}'=0.5$.  
It can indeed be seen that: $i)$ in the  
$Q^2$-rescaling model the ratio is independent of $|\vec P_{A-1}|$, 
$ii)$ when  
$\theta_{\widehat{\vec P_{A-1} \vec q} } =90^o$,  
the $x-$rescaling model predicts a 
$|\vec P_{A-1}|$-independent ratio for $^{40}Ca$ $(z_1^{(40)}\simeq 1)$ 
and a strongly $|\vec P_{A-1}|$-dependent ratio for $D$ 
$(z_1^{(2)}\simeq 1 - \frac{ |\vec P_{A-1}|^2}{2M^2} )$; when  
$\theta_{\widehat{\vec P_{A-1} \vec q} }=180^o$,  
also the ratio for $^{40}Ca$ becomes strongly 
$|\vec P_{A-1}|$-dependent, for, now, $z_1^{(40)}\simeq 1 -  
\frac{ |\vec P_{A-1}|}{M}$. To sum up, it can be seen that the semi-inclusive  
process allows one to choose a variety of kinematical conditions which 
enhance various aspects of the problem.  
 We have seen in  
Section III-B that, due to the small values 
of $|\vec {P_{A-1}}|$ and $E_{min}$, 
the process $A(e,e'(A-1))X$ on a complex nucleus is characterized by 
$z_1^{(A)}\simeq 1$ when  
$\theta_{\widehat {\vec P_{A-1} \vec q}}=90^o$; 
as a result,
the off-mass-shell dependence of $F_2^{N/A}$ 
disappears (cf. full curve in Fig. \ref{fig5}); 
the off--mass--shell dependence of $F_2^{N/A}$ can on the contrary be enhanced 
if $\theta_{\widehat{ 
\vec P_{A-1} \vec q}}=180^o$,  
for an appreciable contribution from the last term of Eq. (\ref{zac}) 
is now generated; 
if so, however, the ratio (\ref{x1x2x}) for a complex 
nucleus will not appreciably differ from that of the deuteron  
(cf. full curves  in Fig.\ref{fig6}), since off-mass-shell effects 
are solely due to the nucleon momentum $| \vec P_{A-1}|$, and not to
medium effects provided by e.g. the nucleon binding ($E_{min}/M << 1$). Possible modifications of  $F_2^{N/A}$ due to medium effects, will be discussed in the next Section.
 
\section{The {\bf \it A\lowercase{(e,e'} $N_2$(A-2))X} process}

In the previous Section we have discussed the case of weakly bound,  
non-correlated nucleons.  
In the present section we will investigate the 
semi inclusive processes occurring on a strongly correlated 
nucleon pair. 
 To this end, 
let us consider  
  the process depicted in Fig. 1 (b), which represents 
   the absorption 
of the virtual photon by a correlated nucleon $N_1$ (with high momentum 
$|\vec P_{A-1}|$), followed by the emission 
of the partner nucleon $N_2$ (   
with momentum $\vec p_2$), 
and by the recoil of the  
$(A-2)$ system, 
with low momentum 
$\vec P_{A-2}=-(\vec P_{A-1} + \vec p_2)$ 
and low excitation energy. 
 The experimental investigation of  
such a process would require the coincidence detection of the scattered 
electron, the nucleon $N_2$ and the system $(A-2)$. 
       
The differential cross section of the process reads as follows

   \begin{eqnarray} 
   \sigma_2 ^A(x_{Bj}, & Q^2 &,\vec P_{A-2}, \vec p_2) 
    \equiv   \frac{d\sigma^A}{d x d Q^2 
   d \vec P_{A-2} d \vec p_{2}}  = 
   \nonumber\\[1mm] 
   & &  
    K^A( x_{Bj},Q^2,y_A,z_1^{(A)})  
      z_1^{(A)}  F_2^{N/A}(x_{Bj}/z_1^{(A)},Q^2) 
      n^A_{cm}(|\vec P_{A-2}|)\, n^A_{rel.}(|\vec p_2 +\vec P_{A-2}/2 |)
   \label{crosa-2} 
   \ee 
where $P_{N_1,N_2}$ is the two-nucleon spectral function, defined 
in Section II,  and $K^A$, $y$, $y^A$, $x^A$, and $z_1^A$
  are defined by  
Eqs.  (\ref{ka}), (\ref{ydef}),  (\ref{adef}), and (\ref{zan})with 
\begin{equation} 
\vec p_1 = -\vec P_{A-1} =  (\vec p_2 + \vec P_{A-2})~. 
\label{queva} 
\end{equation} 
and
\begin{equation} 
p_{10}=M_A-\sqrt{M^2+\vec p_2^2} 
-\sqrt{M_{A-2}^2+\vec P_{A-2}^2}. 
\label{aa2} 
\end{equation} 
In the Bjorken limit  $\eta=1$ one has  
\begin{equation} 
z_1^{(A)}=\frac{M_A}{M}-z_2-\frac{M_{A-2}}{M}z_{A-2}, 
\label{aa3} 
\end{equation} 
where  
\begin{equation} 
z_2=\frac{\sqrt{M^2+\vec p_2^2} - |\vec p_2 | 
\cos \theta_{\widehat{\vec p_2 \vec q}}}{M} 
\label{aa4} 
\end{equation} 
and  
\begin{equation} 
z_{A-2}=\frac{\sqrt{M_{A-2}^2+\vec P_{A-2}^2}-|\vec P_{A-2}| 
\cos \theta_{\widehat{\vec P_{A-2}\vec q}}}{M_{A-2}} 
\label{aa5} 
\end{equation} 
are the light cone momentum fraction 
of  nucleon $N_2$ and nucleus $(A-2)$, respectively. 
Note that Eq. (\ref{aa3}) is nothing but the energy 
conservation of the process 
 \begin{equation} 
\nu +M_A=\sqrt{M_x^2+(\vec p_1+\vec q)^2} + \sqrt{M^2+\vec p_2^2} 
+ \sqrt{M_{A-2}^2+\vec P_{A-2}^2} 
 \label{conserv1} 
 \end{equation}  
in the Bjorken limit. 
In the non relativistic approximation one obtains 
 \begin{equation} 
z_1^{(A)} \simeq 1 - {E \over M} - {|\vec P_{A-1}|^2 \over 2 (A-1)M^2} 
+ \frac{|\vec P_{A-1}|}{M}  cos \theta_{\widehat{\vec P_{A-1} \vec q} }~, 
\label{zac21} 
\end{equation} 
with $ E = (E_{th}^{(2)} + E_{A-1}^*)$.  
Due to the small value of $\vec k_{cm} = - \vec P_{A-2}$, we can write 
$E_{A-1}^* \simeq {(A-2) \over 
2M(A-1)} |\vec P_{A-1}|^2$, and by considering that $\vec P_{A-1} = -(\vec p_2 
+ \vec P_{A-2})$ (cf. (\ref{queva}) ), one gets: 
\begin{equation} 
 E = E_{th}^{(2)} + {(A-2) \over 2 M (A-1)}  
\left [  
|\vec p_2|^2 
+ \left (  
{A-1 \over A-2} 
\right )^2  
\vec P_{A-2}^2 + 2 { A-1 \over A-2 } 
|\vec p_2| |\vec P_{A-2}|  
cos \theta_{ \widehat {\vec p_2 \vec P_{A-2}}} \right ] ~. 
\label{30a} 
\end{equation} 
It should be stressed  that    the nucleon structure function $F_2^{N/A}(x_{Bj}/z_1^{(A)},Q^2)$  
  appearing in Eqs. (\ref{crosa-1})  and  (\ref{crosa-2}) 
  reflects different physical situations, for  
   in the first case $F_2^{N/A}$ represents  
   the quark distribution in a weakly bound, quasi-free nucleon, whereas, in 
   the second case,  
it represents the quark distribution in a strongly bound  
nucleon, which, in principle,  can undergo, because of  binding, 
 off-mass-shell deformations 
   (see, for instance refs. \cite{kukhanna,mthomas}).  
   Therefore, if the nucleon structure function could be 
   extracted from the cross section  
   (\ref{crosa-2}) and compared  
   with the one obtained  from  the cross section (\ref{crosa-1}), 
   a direct comparison of nucleon structure functions 
   for weakly bound and deeply bound nucleons could, for the first time, 
   be carried out. 
    
   It should be pointed out that, 
   since $y_A$ depends upon the high momentum $|\vec p_2|$, 
   the factor $ K^A(x_{Bj},Q^2,y_A)$ may strongly 
   differ from $ K^N(x_{Bj},Q^2,y)$, 
   unless one of the two following kinematical conditions  
   are chosen: $i)$ small values of $x_{Bj}$; $ii)$ the Bjorken limit. 
    We found that   
   at  $Q^2=20 GeV^2/c^2 $ and $x_{Bj}=0.05$, 
   the direction of the momentum transfer $\vec q$ 
   coincides, in the frame where the target 
   is at rest, with the electron beam direction 
   ( $ \theta_{\widehat {\vec k \vec q}}\approx 2^0$ ); in this case, 
   $y_A\simeq y$ and $K^A \simeq K^N$ 
   (our numerical estimates 
   show that $K^A/K^N$ 
   varies from $0.99$ at $|\vec p_2|=350$ MeV/c to $0.96$ at  
   $|\vec p_2|=1$ GeV/c); adopting realistic figures  
   for an electron-ion collider, 
   i.e. ${\cal E}_e \approx 5 $ GeV, $T_N$ = (kinetic 
   energy per nucleon) $\approx 25$ GeV \cite{gsi} in its laboratory system, 
   the chosen values of $Q^2$ and $x_{Bj}$ correspond 
   to ${\cal E}_e'\approx 2\,GeV;\quad \theta_{\widehat {\vec k 
   \vec k'}} \approx 90^0$ 
   (in the nucleus rest frame they would 
   correspond to ${\cal E}_e\sim 260$ GeV,  
   $ {\cal E}_{e'}\approx 50$ GeV; $\theta_{\widehat {\vec k \vec k'}}  
\approx 2^0$). 
 
\subsection{Checking the spectator mechanism in the semi-inclusive 
process $A(e,e'N_2(A-2))X$} 
 
   The validity of eq. (\ref{crosa-2}) can  experimentally be tested 
   by taking advantage of the observation ~\cite{pan} that for high  
   values of $|\vec P_{A-1}|$ the nucleon 
   momentum distribution for a complex nucleus turns out 
   to be the rescaled momentum distribution of the deuteron, 
   with very small $A$ dependence (unlike what happens 
   for the low momentum part of $n(k)$ (cf. Fig. 2)). Let us 
   therefore consider the following ratio, where $|\vec P_{A-2}|=|\vec P_{A'-2}|$: 
   \begin{eqnarray} 
   R(x_{Bj},Q^2,\vec P_{A-2},\vec p_2,z_1^{(A)},z_1^{(A')} )\, & \equiv\,&  
   \displaystyle\frac{\sigma_2^A(x_{Bj},Q^2,\vec P_{A-2},|\vec p_2|)} 
   {\sigma_2^{A'}(x_{Bj},Q^2,\vec P_{A-2},|\vec p_2|)} 
   \nonumber\\[1mm] 
   & = &  
   \frac{z_1^{(A)}}{z_1^{(A')}} 
   \frac{F_2^{N/A}(x_{Bj}/z_1^{(A)},Q^2)}{F_2^{N/A'}(x_{Bj}/z_1^{(A')},Q^2)} 
   \cdot 
   \frac{n^A_{rel.}(|\vec p_{rel.}|)}{n^{A'}_{rel.}(|\vec p_{rel.}|)} 
   \cdot 
   \frac{n^A_{cm}({ |\vec P_{A-2}| })}{n^{A'}_{cm}({ |\vec P_{A-2}| })}~, 
   \label{ratio2} 
   \end{eqnarray}  
   where $ |\vec p_{rel}| =|\vec p_2 +\vec P_{A-2}/2 |$. 
   If $|\vec P_{A-2}|$ is fixed, then, provided 
   $F_2^{N/A} = F_2^{N/A'}$, the ratio, 
   measured at $p_{rel} \geq  2 - 3$ $fm^{-1}$, would be roughly a constant, 
   since $n^A_{rel.}\propto 
   n^{D}$ for any $A$. 
   The condition $F_2^{N/A} = F_2^{N/A'}$ can be achieved  
    by properly  
   choosing, for $A$ and $A'$, the values of 
   $\vec p_2$ and $\vec P_{A-2}$  
    appearing in (\ref{30a}), so as to make  
   $z_1^{(A)} \simeq z_1^{(A')}$, i.e. 
   $F_2^{N/A} \simeq F_2^{N/A'}$ (note, moreover, that for large values 
   of $| \vec P_{A-1}|$ and large values of $A$, 
   the dependence of $z_1^{(A)}$ upon $A$ is unessential).  
To summarize, the cross-section  
(\ref{crosa-2}) should be measured  
for the systems $A$ and $A'$ at the same values of  
$x_{Bj}$, $Q^2$ and $\vec P_{A-2}$, changing the values of the angle  
$\theta_{\widehat {\vec p_2 \vec P_{A-2} }}$ and $|\vec p_2|$ so as to vary  
$|\vec p_{rel}| = |\vec p_2 + \vec P_{A-2}/2|$, keeping $z_1^{(A)}=z_1^{(A')}$. 
If Eq. (\ref{crosa-2}) is basically correct, the ratio 
(\ref{ratio2}) plotted versus $|\vec p _{rel}| \geq 2 - 3$ fm$^{-1}$ 
should exhibit (as shown in Fig. \ref{fig7})the same deuteron-like behaviour 
for any two nuclei in the range, say, 
$2<A<208$. If  such a deuteron -- like 	behaviour of eq. (\ref{ratio2})  
    is experimentally found, it would represent a stringent test 
    of the spectator mechanism. A word of caution is however in order here: the FSI between the nucleon $N_2$ and the nucleus $(A-2)$ will presumably affect the ratio  (\ref{ratio2}). Calculations of the FSI within the Glauber multiple scattering approach are in progress and will be reported elsewhere; preliminary results indicate that in the region of the considered kinematics, the replacement of the undistorted two-body Spectral Function with the distorted one, mainly affects the absolute value of the ratio  (\ref{ratio2}).

\subsection{Investigating the structure functions of deeply bound  
nucleons by the process $A(e,e'N_2(A-2))X$} 
 
As in the case of the $A(e,e'(A-1))X$ process, in order to investigate 
the structure function of a (deeply) bound nucleon, we have to figure 
out experimentally measurable  
quantities which could provide information on $F_2^{N/A}$ without 
contaminations from the nucleon momentum distributions, or other 
momentum dependent terms. To this end, we use the  
analog of the ratio (33) 
which, within the convolution model, assumes the following form 
 
   \begin{eqnarray} 
   R(x_{Bj},x_{Bj}',Q^2,|\vec P_{A-2}|,|\vec p_2)|\, & \equiv\,&  
   \displaystyle\frac{\sigma_2^A(x_{Bj},Q^2,|\vec P_{A-2}|,|\vec p_2|)} 
   {\sigma_2^{A}(x_{Bj}',Q^2,|\vec P_{A-2}|,|\vec p_2|)} 
   \nonumber\\[1mm] 
   & = & \frac{x_{Bj}'}{x_{Bj}}  
   \frac{F_2^{N/A}(x_{Bj}/z_1^{(A)},Q^2)}{F_2^{N/A}(x_{Bj}'/z_1^{(A)},Q^2)}~. 
   \label{ratios} 
   \end{eqnarray} 
 
It should be pointed out that, although the r.h.s. of eqs. (\ref{x1x2x}) 
and (\ref{ratios}) look formally the same, they differently  
depend upon the nucleon binding, for, we reiterate, in Eq. (\ref{x1x2x})  
one has (cf. Eq. (\ref{zac})) 
 
\begin{equation} 
z_1^{(A)} \simeq 1 - {E_{min} \over M} - {|\vec P_{A-1}|^2 \over 2 (A-1)M^2} 
+ \frac{|\vec P_{A-1}|}{M}  cos \theta_{\widehat {\vec P_{A-1} \vec q }}~. 
\label{zacn} 
\end{equation} 
 with $\vec P_{A-1} \equiv -\vec {p_1}$, whereas in Eq. (\ref{ratios}), one has 
(cf.  Eq. (\ref{zacn}) ) 
 
\begin{equation} 
z_1^{(A)} \simeq 1 - {E \over M} - {|\vec P_{A-1}|^2 \over 2 (A-1)M^2} 
- \frac{|\vec P_{A-1}|}{M}  cos \theta_{\widehat {\vec p_1 \vec q}}~, 
\label{zac2} 
\end{equation} 
with $\vec P_{A-1} = -(\vec p_2 + \vec {P_{A-2}})$ 
and $E$ given by Eq. (\ref{30a}). 
We have seen in  
Section 3.2 that, due to the small values 
of $|\vec {P_{A-1}}|$ and $E_{min}$, 
the process $A(e,e'(A-1))X$ on a complex nucleus is characterized by 
$z_1^{(A)}\simeq 1$, when  
$\theta_{\widehat {\vec P_{A-1} \vec q}}=90^o$, 
with the result that  
the off-mass-shell dependence of $F_2^{N/A}$ 
disappears (cf. the full line for $^{40}Ca$  in Fig.\ref{fig5}); 
the off--mass--shell dependence of $F_2^{N/A}$ can be enhanced 
if $\theta_{\widehat{ 
\vec P_{A-1} \vec q}}=180^o$,  
for an appreciable contribution from the last term of Eq. (\ref{zac2}) 
is generated; 
if so, however, the ratio (\ref{x1x2x}) for a complex 
nucleus will not appreciably differ from that of the deuteron  
(cf. the full curves in  in Fig.\ref{fig6}), since off-mass-shell effects 
are solely due to the nucleon momentum $| \vec P_{A-1}|$, with no contribution  
from nucleon binding ($E_{min}/M << 1$); 
 a totally different 
situation is expected to occur in the process $A(e,e'N_2(A-2))X$; 
as a matter of fact, in this case the
 ``binding term'' ${  E/M}$ in Eq. (\ref{zac2}) 
will generate an appreciable contribution to $z_1^{(A)}$, 
 due to the large value of $|\vec p_2|$ associated  to nucleon-nucleon 
correlations  
(or, equivalently, to high values 
of the removal energy $E$). Thus, in order to check whether the structure for  
a deeply bound nucleon would dynamically differ from the one for a weakly 
bound one, the ratios (\ref{x1x2x}) and (\ref{ratios}) 
for a given nucleus should be plotted versus the same value of $z_1^{(A)}$; 
in such a way, the  
off-mass-shell dependence of $F_2^{N/A}$ is quantitatively the same, 
but it originates 
from different contributions to $z_1^{(A)}$, viz. the momentum $\vec P_{A-1}$, 
for the weakly bound nucleon, and the binding effect $E$, for 
a deeply bound nucleon. If a different behaviour of the 
two ratios is found,  
this would represent strong evidence that the structure functions for 
weakly and deeply bound nucleons are different. Here, again, the $N_2-(A-2)$
FSI should be taken into account, although its effect is expected to be canceled in the ratio (\ref{ratios}).
 
    Another possibility to investigate the nucleon 
structure functions would be to analyze the following ratio 
 
\begin{equation} 
R_2=\frac { \sigma_2 ^A } { \sigma_1^D } 
   =\frac{ z_1^{(A)}\, F_2^{N/A}  
%( x_{Bj} /z_1^{(A)}, \xi_A (Q^2) Q^2 ) 
( x_A, Q^2, p_1^2 )\, 
          n_{cm}^A( |\vec P_{A-2}| )\,n_{rel.}^A (|\vec p_2 + \vec P_{A-2}|)} 
         {z_1^{(D)} \,F_2^{N/D} 
%( x_{Bj}/z_D, \xi_D(Q^2)Q^2 )  
( x_D,  Q^2, p_1^2 )  
          \, n^D ( |\vec P_{A-1}| = |\vec p_2 + \vec P_{A-2}| ) } 
\label{ratio3}~. 
\end{equation} 
 \\ 
\noindent 
The results  for $A=12$ are presented in Fig.\ref{fig8}, 
 in correspondence  
of two values of $|\vec p_2|$, for fixed values of the following kinematical variables:  
   $|\vec P_{A-2}| = 50 MeV/c$, $\theta_{  \widehat{p_2q}} = 90^0$, 
     $\theta_{  \widehat{p_2P_{A-2}}} = 180^0$, and 
         $Q^2=20 GeV^2/c^2$; 
the full and dashed curves refer to the $x$- and $Q^2$-rescaling 
models, respectively. Let us first analyze the results predicted by the first model, viz. $F_2^{N/A} (x_A,Q^2,p_1^2) \rightarrow  
F_2^{N/A} (x_{Bj}/z_1^{(A)},Q^2,p_1^2)$ 
with $z_1^{(A)}$ given by (\ref{zac21}).  
 The value of the three-momentum of  
    the $(A-1)$ fragment (a nucleon) in the $A(e,e'(A-1))X$ cross section 
    off the deuteron, has been chosen the same as the 
    three-momentum of the interacting nucleon $\vec P_{A-1}$ in the 
    case of the $A(e,e',N_2(A-2))X$ process off $^{12}C$; by this choice,
 the removal energy which appears in $z_1^{(A)}$  
(\ref{zac2}) is almost 
equal to the recoil energy appearing in $z_D$, 
so that $z_1^{(A)} \simeq z_1^{D}$; by this way one should expect 
a constant behaviour of $R_2$  (note that $K_A \simeq K_D$, 
for in both cases one has to do with the same values of  
$\vec P_{A-1}$);  the deviation from 
a constant exhibited by the full lines in Fig.\ref{fig8}  
is due to the fact that, with the chosen kinematics, 
$z_1^{(D)} > z_1^{(A)}$. Again, the observation of a behaviour different from  
the one  
presented in Fig.\ref{fig8} would indicate a dependence  
of $F_2^{N/A}$ upon the 
binding energy. Let us now consider the prediction by the  
$Q^2$-rescaling model. For the latter, we have considered the model of Ref.  
\cite{mthomas}, where the renormalization scale associated  
to the momentum of a bound nucleon is given by its invariant mass,  
$p_1^2  \not= M^2$ . Such an assumption leads to the ansatz 
$F_2^{N/A}(x_A,Q^2,p_1^2)= F_2^N(x_A,\xi_A(Q^2,p_1^2)Q^2)$  
with 
$\xi_A(Q^2,p_1^2)= 
 { (M^2/p_1^2) } ^{(\alpha(p_1^2))/ (\alpha(Q^2))}$, i. e. to    
an explicit dependence  
upon the off--shellness of the nucleon .  
Since the invariant mass 
of a deeply bound nucleon strongly differs from $M^2$, 
the ratio (\ref{ratio3}) gets the strong  
$x_{Bj}$ and $|\vec p_2|$ 
dependence shown in Fig. 8. 
 
\section{The local $EMC$ effect} 
 
In the binding model ($x-$ rescaling) of the $EMC$ effect,  
the slope of the ratio  
$R(x_{Bj},Q^2)$=$F_2^A(x_{Bj},Q^2)/(A F_2^N(x_{Bj},Q^2))$ 
 is generated 
by the   average value of the nucleon removal energy $<E>$: 
 the larger the value of $<E>$, the stronger the $EMC$ 
effect\cite{fs}. Since $NN$ correlations produce high values of $E$, 
and therefore strongly affect the value of $R$ \cite{simo},  
it would be extremely interesting to measure the so-called {\it local 
EMC}  (LEMC) effect, i.e., the separate contribution to the 
ratio $R$ of the weakly   and deeply bound nucleons. Several calculations 
of the local $EMC$ effect  
appeared \cite{loc1,loc2}, and attempts have also been made to  
compare them with experimental data on neutrino-nucleus  
DIS \cite{loc3}, but the comparison was not conclusive 
due to the apparently very large   contaminations 
of the 
data from non nuclear effects, like e.g. 
quark fragmentation.   
 
The  semi-inclusive $A(e,e'(A-1))X$ and $A(e,e',N_2(A-2))X$ processes, 
offer the possibility to investigate the LEMC effect. As a matter 
of fact, let us consider 
  the cross sections (\ref{crosa-1}) 
and (\ref{crosa-2}) for a nucleus $A$ and the cross section 
(\ref{crosa-1}) for the deuteron, integrated over a certain interval 
of $\vec P_{A-1}$, with  $\vec P_{A-1} = - 
\vec p_1$ in (\ref{crosa-1}),  and $\vec P_{A-1} = - 
( \vec p_2 + \vec P_{A-1} )$,  in (\ref{crosa-2}).  
The following two quantities  
\begin{equation} 
R_0(x_{Bj},Q^2) = {  
\int_a^b \sigma_1^A( x_{Bj},Q^2,\vec P_{A-1})d \vec P_{A-1} 
\over 
\int_a^b \sigma_1^D( x_{Bj},Q^2,\vec P_{A-1}) 
 d \vec P_{A-1}}~, 
\label{r0} 
\end{equation} 
 
\begin{equation} 
R_1(x_{Bj},Q^2) = {  
\int_a^b \sigma_2^A( x_{Bj},Q^2,\vec P_{A-2}, \vec p_2) 
d \vec P_{A-2} d \vec p_2 
\over 
\int_a^b \sigma_1^D( x_{Bj},Q^2,\vec P_{A-1}) 
d \vec P_{A-1}}
\label{r1} 
\end{equation} 
will therefore provide the LEMC effect, for they 
represent the contribution from weakly bound (\ref{r0}) and 
strongly bound (\ref{r1}) nucleons, respectively \cite{loc2}.   
Since the calculation of 
Eq. (\ref{r1}) is a bit involved, we will consider a more 
restricted type of LEMC, namely   the separate contributions 
of the  EMC effct from the various shells of a 
complex nucleus, i.e.  
the separate contribution of the various shells to the ratio $R_0$ 
\cite{loc1}. 
This means that we will assume that the 
energy resolution in the process  $A(e,e'(A-1))X$ is such,  
that the contribution to the ratio $R_0$ due to  
the ground state, and to the excited states corresponding to the  
hole state of the target, can experimentally be separated. 
In what follows    the $^{12}C$ nucleus will be considered assuming that DIS 
occured on a neutron; this 
 means that the final nucleus  to be detected  
 is $^{11}C$ in the ground state (deep inelastic scattering on a $p$-shell neutron) and in  
 an excited state with excitation  
energy of about $20MeV$  
(deep inelastic scattering on a $s$-shell neutron).  
We have therefore calculated 
the  ratio(\ref{r0}) using realistic Hartree-Fock momentum 
distributions for the $s$ and $p$ shells  with single-particle 
energies $\epsilon_{0s}$= 36 $MeV$ and $\epsilon_{0p}$= 16 $MeV$. The results are presented in   
Fig.\ref{fig9}, where the usual inclusive EMC ratio, i.e.  
Eq.(\ref{r0}) integrated over the full space, is compared  with  
the separate contribution 
from the $s$ and $p$ shells; it can be seen that, in agreement  
with  \cite{loc2},   the $s$ shell exhibits a stronger EMC effect, but since in $^{12}C$ there are 4 $s$ shell and 8 $p$-shell 
nucleons, the total EMC effect   is  less.  
In what follows, we will consider the ratio $R_0$ integrated in a  
restricted space, viz $0<{|\vec P_{A-1}|}<2$ ${fm}^{-1}$ and  
$0{^o}<\theta_{\widehat {\vec P_{A-1}\vec q}}  
<20^o$  (the nucleus (A-1) is emitted forward) and  
$160{^o}<\theta_{\widehat {\vec P_{A-1}\vec q}}  
<180^o$ (the nucleus (A-1) is emitted backward); in  Fig.\ref{fig10} 
the forward and backward ratios  
are compared with the full inclusive ratio, and it can be seen that  
the latter results from the sum of two almost equal contributions.  
In what follows, only 
 backward emission will be considered, for  this is 
 expected to be less affected by FSI between the (A-1) nucleus and  
the hadrons resulting from quark hadronisation. The semi-inclusive  
backward ratio is shown in  Fig.\ref{fig11} together with  
the separate contributions  
from the $s$ and $p$ shells; it can be seen that not only  the shell 
contributions are well separated, but that  the LEMC effect is much 
larger than the usual EMC effect.    
In order to give a flavor of the order of magnitude of the  
cross sections involved, these are presented in Fig. \ref{fig12} . 
 
\section{Summary and Conclusions} 
 
In the present paper, two new types of semi-inclusive DIS processes of  
leptons off complex nuclei, have been investigated. The first one,  
the process $A(e,e'(A-1))X$ , represents DIS on a shell model,  
low momentum and low removal energy nucleon, followed by the coherent, low 
momentum recoil, of the spectator nucleus $(A-1)$ in the ground, or in a   
low energy excited state; the second one, the processs  
$A(e,e'N_2 (A-2))X$, represents DIS on a nucleon $N_1$ 
of a correlated pair, followed by the emission of the 
 high momentum nucleon $N_2$ 
of the pair, 
and the low momentum spectator nucleus $(A-2)$ in the ground, or in a low energy excited  state.  
The experimental investigation of these processes 
would imply  
the coincidence detection of $e'$ and $(A-1)$, in the first case,  and  
$e'$, $N_2$ and $(A-2)$, in the second case, respectively .  
We have demonstrated that both processes 
can provide relevant information on the following topics:  
 
$i)$ the relevance and nature of the FSI between the  
hadronic jet with the nuclear medium;  
 
$ii)$ the validity of the spectator model; 
 
$iii)$ the  off-shell deformation of the nucleon structure function  
in the nuclear medium and the A-dependence of the ratio of the 
n/p structure functions;  
 
$iv)$ the origin of the EMC effect.  
 
As a matter of facts :  
 
$i)$ if nuclei $(A-1)$ and  $(A-2)$ are detected  
in coincidence with the scattered electron,this is a clear signal of the absence of FSI;
 at the same time, the amount of observed nuclei, i.e. the cross section, will of course 
depend upon the  FSI,  therefore the investigation of its absolute value and its dependence upon  $A$, would allow one to investigate the nature of the FSI, e.g. the hadronisation lenght of the hit quark; 
 
$ii)$ by a proper choice  of the kinematics, the ratio of  the cross section  
$ \sigma [A(e,e'(A-1))X]$ to the cross section  $ \sigma  [D(e,e'N)X]$,  
measured  versus 
$|\vec P_{A-1}|$ = $|\vec p_N| \equiv |\vec p|$, at a fixed  
value of the Bjorken scaling variable $x_{Bj}$, has been shown  
to depend , within the Spectator model approach, only upon the low  
momentum part of the nucleon momentum  
distributions $n_A(|\vec p|)$ and $n_D(|\vec p|)$, and since 
these sharply differ for $|\vec p|\leq 1 fm^{-1}$ , the ratio  
should exhibit a strong $|\vec p|$ dependence (cf.  Fig. \ref{fig2}),  
whose experimental observation would represent a stringent check of  
the validity of the spectator model. At the same time, the ratio of   
the cross section 
$ \sigma [A(e,e'N_2 (A-2))X]$ to the cross section  
$ \sigma [D(e,e'N)X]$ measured  versus  
$|\vec p_{rel}| =|\vec p_2 +\vec P_{A-2}/2 |$  for fixed value of  
  $|\vec P_{A-2}|$ and fixed value of  $x_{Bj}$ , has been  
shown to depend  only upon the relative  momentum distributions  
$n^A_{rel}(|\vec p_{rel}|)$ and $n^D(|\vec p_{rel}|)$, so that 
the ratio should exhibit a $|\vec p_{rel}|$ dependence  
similar for all values of  A, for $n^A_{rel} \sim n^D$  
for $|\vec p_{rel}| \geq 2 fm^{-1}$ (cf. Fig. 7); again, the experimental  
observation of such a scaling behaviour would also represent a  
stringent test of the Spectator model mechanism; 
 
$iii)$ it has been shown that by a proper choice of the kinematics, the ratio of  
the cross sections for the same nucleus but at two different  
values of $x_{Bj}$, becomes  
   independent of the nuclear quantities, being determined only  
by the nucleon structure function; it has therefore been demonstrated, in the case of the process  $A(e,e'(A-1))X$, that such a ratio  could provide significant information  on different models of the structure function of weakly bound nucleons (cf. Figs. 5 and 6). Eventually (cf. Fig. 8) it has
  been shown that the ratio of the cross section  
for the process  
$A(e,e'N_2 (A-2))X$ to the deuteron cross section, could provide  
information on the  {\it binding energy dependence} of the nucleon structure functions;
 
$iv)$ the local EMC effect has been investigated (cf. Figs. 9-12), pointing out that that  
the processes 
$A(e,e'(A-1))X$ and $A(e,e'N_2(A-2))X$  integrated over a proper  
value of the momenta of the detected particles ( $A-1$, $N_2$ and $A-2$)  
will provide, 
for the first time, the separate contribution to the EMC ratio of the  
weakly and deeply bound nucleons, thus providing 
a stringent check of the binding model (x-rescaling) of the EMC effect. 
Detailed calculations have been performed for the process $A(e,e'(A-1))X$,
demonstrating that in the binding model, the inclusive $EMC$ effect 
 results from the cancellation of two large  
contributions from the forward and backward emitted $(A-1)$ nuclei (cf. Fig. 10); therefore,
a significant check of the binding model could  be provided by the measurement 
of the  backward ratio which exhibits a 60 percent deviation from  
unity instead of the  
10 percent deviation of the usual inclusive EMC effect (cf. Fig. 11). 
 
In closing, we would like to point out  that the results we have exhibited  
have been obtained with non relativistic momentum distributions and spectral  
functions. Calculations for the two- and three-body systems including  
relativistic effects by a full covariant Bethe-Salpeter  
approach and by   light-cone spectral functions,
respectively, will be presented elsewhere ( \cite{bssemi}, \cite{lcsemi}).

\vskip 6mm 
\leftline{\Large{\bf Acknowledgements}} 
\vskip 4mm 
\noindent 
We gratefully acknowledge Boris Kopeliovitch and Daniele Treleani for useful comments and discussions. L.P.K. thanks  INFN Sezione di Perugia for 
  warm hospitality and financial support.    
S.S. thanks the TMR programme of the European Commission 
ERB FMR-CT96-008, INFN Sezione di Perugia 
and  Department of Physics, University of Perugia 
for partial financial support.

\appendix 
\section{The electron-hadron cross section}

\indent  
 
In this Appendix the derivation of the semi-inclusive
electron-hadron 
cross section within the instant-form dynamics will be presented.

In the one-photon exchange approximation
the cross section describing the scattering of an  electron $e$ from a hadron $A$ 
 reads as follows: 
\begin{equation}\!\!\!\!\! 
d\sigma= 
\frac{M_A m_e}{(P_A \cdot k_e)}(2\pi)^4\delta^{(4)}(P_A+k_e-k_{e'}-P_f) 
\left | \langle k_{e'}|\hat j^\mu(0)|k_e\rangle\frac{1}{Q^2} 
 \langle P_A|\hat J_\mu^A(0)|P_f\rangle\right |^2\frac{m_ed^3k_{e'}}{{\cal E}_{k'}(2\pi)^3}d\tau_f, 
\label{app1} 
\end{equation} 
where $\hat j^\mu(0)$ and  $\hat J_\mu^A(0)$ are the  
electromagnetic current operators for the lepton and the hadron, respectively, 
$M_A$  ($P_A$, $E_A$), and $m_e$  ($k_e$, ${\cal E}_{k'}$) stand for the masses (4-momenta, total energy) of the hadron and the electron in the initial state,    
 $k_{e'}$ and $P_f$ denote the four-momenta of the electron and the hadron in the final state,  $Q^2 =-q^2= -(k_e-k_{e'})^2 = \vec q^{\,\,2} - \nu^2=4 {\cal E}_k  
{\cal E}_{k'} sin^2 {\theta \over 2}$ is the 4-momentum transfer,  
(with $\vec q = \vec k_e - \vec k_{e'}$, $\nu= {\cal E}_k - 
{\cal E}_{k'} $ and $ \theta \equiv \theta_{\widehat{\vec k_e  \vec k_{e'}}}$),  and  
$d\tau_f$   the   phase space volume of all particles (but the scattered electron)
in the final state. 
In Eq. (\ref{app1}) the following normalization conditions are 
used: 
\begin{equation} 
\Lambda_+(p) = {\hat p+M \over 2M},\quad 
\langle p | p' \rangle = {E \over M} (2 \pi)^3 \delta^3(\vec p -\vec p'), 
\quad \bar u\,u = 1,\quad u^+u=\frac{E}{M}. 
\label{norm} 
\end{equation} 
where $M$ is the nucleon mass.

\subsection{The inclusive cross-section} 
 
By placing in Eq. (\ref {app1}) $f \equiv X$ and $d\tau_f = 1$, in the lab system, the 
 cross section  for the inclusive process $A(e,e')X$, i. e. when only the scattered electron is detected, is obtained 
\be 
{d \sigma \over d \Omega ' d {\cal E}_{k'}} = 
{4 \alpha^2 \over Q^4} {{\cal E}_{k'} \over {\cal E}_k}  
{1 \over 2} L^{\mu\nu} W^A_{\mu\nu} 
\ee 
where the leptonic tensor,  $L_{\mu\nu}$, is 
\be 
L_{\mu\nu} = k_{\mu} k_{\nu}' + k_{\mu}' k_{\nu} - 
g_{\mu \nu} (k \cdot k'), 
\label{ll} 
\ee 
and the hadronic tensor, $W_{\mu\nu}^A$, is
 \be 
 W_{\mu\nu}^A&=&  { 1 \over 4 \pi}\overline{\sum_{\alpha_A}}  \sum_X   
(2\pi)^4 \delta^{(4)} (P_A + q - p_X)\nonumber\\ 
& &
\langle \alpha_A, \vec P_A=0 | J_\mu^A(0) |  
\alpha_X\vec {p}_X  \rangle \langle  \alpha_X\vec{p}_X|J_\nu^A (0) | 
\alpha_A,\vec P_A=0 \rangle .
\label{nuctz} 
\ee 
The general form of the hadronic tensor,
 restricted by requirements of gauge-invariance, time-reversal invariance and
parity conservation, 
depends upon two  structure functions $W_{i}^A$, corresponding to the two 
independent scalars of the problem, viz.
 \be 
    W_{\mu \nu}^{A} =  W_1^{A}(\nu, Q^2) \left [ g_{\mu \nu} + 
    {q_{\mu} q_{\nu} \over Q^2} \right ] + {W_2^{A} (\nu, Q^2) \over M^2} 
    \tilde{P^A}_{\mu} ~ \tilde{P^A}_{\nu} 
    \label{inc} 
 \ee 
\noindent 
where $\tilde{P^A}_\mu = p_{\mu}^A +  
\displaystyle\frac{q_\mu(p^A \cdot q)}{  Q^2}$.

The contraction of the two tensors gives the  well known result: 
 
\be 
{d \sigma \over d \Omega ' d {\cal E}_{k'}} = \sigma_{Mott} \left [ W_{2}^A(\nu,Q^2) + 2 
W_{1}^A(\nu,Q^2) \tan^2 {\theta \over 2} \right ] 
\ee 
where 
\be 
\sigma_{Mott} = \left ( { \alpha \,\,\cos {\theta \over 2} \over 
2 {\cal E}_k \sin^2 {\theta \over 2} } \right )^2 
\ee  
is the Mott cross section. Note that  the inclusive process on the nucleon $N(e,e')X$, is described by the above formulae with $A=N$, $P^A=p^N$. 

\subsection{The semi-inclusive cross-section}
 
 Let us now discuss the semi-inclusive process 
of the  
type  $A(e,e'B)X$, when another hadron  $B$ is detected in coincidence with the electron.  
We  have in this case $f \equiv(B,X)$ and $d\tau_f = \frac{M_Bd^3P_B}{{ E}_{B}(2\pi)^3}$. The relevant hadronic four-momenta involved in the process are $P_B \equiv {(P_B^0, \vec P_B)}$, with $P_B^0 = \sqrt {(M_B +E_B^*)^2
+ {\vec P_B}^2}$, $M_B$ and $E_B^*$ being, respectively, the rest mass and the  intrinsic excitation energy of $B$, and $p_X \equiv {(p_X^0,\vec p_X)}$, with
$p_X^0 = \sqrt{M_X^2 + {\vec p_X}^2}$.
The cross-section in IA is given by

\be 
{d^4 \sigma \over d \Omega ' d {\cal E}_{k'} ~ dE_B ~ d\Omega_B} = 
{4 \alpha^2 \over Q^4} {{\cal E}_{k'} \over {\cal E}_k}  {{|\vec P_B|}{E_B}  \over {M}_B} 
 L^{\mu\nu} W^{A, s.i.}_{\mu\nu} 
\ee 
where the form of the leptonic tensor is again given by (\ref{ll}),
but the hadronic tensor will have a more complex structure,
viz:

\begin{eqnarray} 
W_{\mu\nu}^{A, s.i.} & = & { 1 \over 4 \pi} {\overline {\sum_{\alpha_A}}} \sum_{ \alpha_B, X}  
(2\pi)^4 \delta^{(4)} (P_A + q - P_B -p_X)
\nonumber
\\ 
& & 
\langle \alpha_A \vec P_A=0| J_\mu^A(0) |  
\alpha_X\vec p_X , \alpha_B\vec P_B  E_B^* \rangle 
\langle \alpha_B\vec P_B  E_B^* ,\alpha_X \vec p_X  | J_\nu^A(0) | 
\alpha_A\vec P_A = 0 \rangle ~, 
\label{hadrontz}
\end{eqnarray} 
where the sum over $X$ stands for a sum over the discrete and and an integral over the continuum quantum numbers 
of $X$,  $\alpha_{B}$   stands for the  discrete and continuum
quantum numbers of the  final nucleus,
and the vector $|\alpha_X \vec p_X, \alpha _B \vec P_B  E_B^* \rangle$ 
consists asymptotically of a nucleus $B$ detected 
with momentum $\vec P_B$ and intrinsic excitation
energy $E_B^*$,
and an undetected hadronic state $X$.
For the semi-inclusive process we are considering,
the general form of the hadronic tensor,
 restricted by requirements of gauge-invariance, time-reversal invariance and
parity conservation, 
depends on four  structure functions $W_{i}^A$ , corresponding to the four 
independent scalars of the problem, viz. (see e.g. Refs. \cite{boffi}and 
\cite{forest} and references therein quoted) 
\be 
W_{\mu\nu}^{A, s.i.} = -{W_{1}^A} g_{\mu\nu} + {{W_{2}^A} \over M^2} P^A_{\mu}P^A_{\nu} 
+ {W_{3}^A} { 1 \over (p \cdot P^A)}\, { 1 \over 2}\, (P^A_\mu p'_\nu + 
P^A_\nu p'_\mu) + {{W_{4}^A} \over M^2} p'_\mu p'_\nu 
\label{s-i} 
\ee   
where the terms  linear in $q^\mu$ do not appear thanks to the gauge invariance  
of  the leptonic tensor. 
The structure of $W_{\mu\nu}^{A, s.i.}$ can
 be obtained in a more physically transparent way, by introducing , instead of $W_{1-4}$, another set   
of four scalar  response functions. 
To this end,  the complete set of polarization 4-vectors 
for a  
virtual photon 
\be 
\epsilon_{\pm}^\mu = \mp { 1 \over \sqrt{2} } (0,1,\pm i,0),\quad 
\epsilon_{0}^\mu =  { 1 \over \sqrt{Q^2}} (|\vec q|,0,0,q_0) 
\label{l} 
\ee 
is introduced, with $\epsilon_\mu q^\mu = 0$, $\sum_\lambda \epsilon_\lambda^{*\mu}  
\epsilon_\lambda^\nu = - g^{\mu \nu} + { q^\mu q^\nu \over q^2}$, and  
$\epsilon_\lambda^* = (-1)^\lambda \epsilon_{-\lambda}$ ($\lambda = \pm,0$), 
to obtain 
\be 
L^{\mu\nu} W_{\mu\nu}^{A, s.i.} = \sum_{\lambda \lambda'} L_{\lambda \lambda'} 
W_{\lambda \lambda'}, 
\label{cont} 
\ee 
where 
 
\be 
L_{\lambda \lambda'} = \epsilon_\lambda^\mu L_{\mu \nu}  
\epsilon_{\lambda'}^{*,\nu}, \quad 
W_{\lambda \lambda'} = (-1)^{\lambda + \lambda'} 
\epsilon_\lambda^{*,\mu} W_{\mu \nu}^{A,s.i.} \epsilon_{\lambda'}^{\nu}. 
\label{wll1} 
\ee 
Due to time-reversal and parity invariances of the electromagnetic 
interaction,  only four 
independent combinations of $\lambda\lambda'$ will appear in  
 Eq. (\ref{cont}), which are usually chosen  
in the following form: 
\be 
\begin{array}{ccc} 
W_{L}^A = \displaystyle\frac{|{\vec q}|^2}{Q^2}\,W_{00}~; && 
W_{T}^A= W_{11}+W_{-1-1}~;\\ 
W_{LT}^A=\displaystyle\frac{|{\vec q}|}{\sqrt{Q^2}}\, 2{\rm Re} \left [ 
W_{01}-W_{0-1}\right ]~;&& 
W_{TT}^A = -2 {\rm Re}\,  W_{1-1}~; 
\end{array} 
\label{response} 
\ee 
defining, respectively, the longitudinal ($L$), transverse ($T$),
longitudinal-transverse ($LT$) interference  and transverse-transverse 
($TT$) nuclear response functions. 
The corresponding parts of the  
leptonic tensor can be straightforwardly found 
by subtracting  a factor   $4{\cal EE}'\cos^2\theta/2$ from 
$L_{\lambda\lambda'}$, eq.
(\ref{wll1} ),
i.e. by defining 
  a ``reduced'' leptonic tensor  
$L_{\lambda\lambda'} = 4{\cal EE}'\cos^2\theta/2\, l_{\lambda\lambda'}$.
One finds 
\be 
\begin{array}{ccc} 
l_{00} = \displaystyle\frac{Q^2} {|{\vec q}|^2}~; && 
l_{11} =\displaystyle\frac{Q^2}{2 |{\vec q}|^2} +  
              \tan^2\displaystyle\frac{\theta}{2}~; \\[2mm] 
l_{01} =\displaystyle\frac{1}{\sqrt{2}} \displaystyle\frac{\sqrt{Q^2}} 
{|{\vec q}|} 
\left ( 
\displaystyle\frac{Q^2}{|{\vec q}|^2}+\tan^2\displaystyle\frac{\theta}{2} 
\right)^{1/2}~;  && 
l_{1-1}=-\displaystyle\frac{Q^2}{2|{\vec q}|^2} 
\end{array} 
\label{lept} 
\ee 
 and the cross section assumes the well-known form (see e.g. \cite{forest} who cosidered the process $A(e,e'p)X$ in the quasi-elastic region): 
\be 
    {d^4 \sigma \over  d \Omega ' d {\cal E}_{k'} ~ dE_B ~ d\Omega_B} =  
    \sigma_{Mott} ~ |\vec p_B| ~ E_B ~ \sum_i ~ V_i ~ W_{i}^A( \nu , Q^2, \vec{p}_B, E_B^f) 
     \label{1}  
 \ee 
where 
$i \equiv\{L, T, LT, TT\}$,
and the kinematical factors  $V_i$ , in agreement   with 
the definitions (\ref{response}),   
(\ref{lept}), have the following form: 
\be 
\begin{array}{ccc} 
 V_L = \displaystyle\frac{Q^4}{ |{\vec q}|^4}~,&&  
 V_T = \tan^2(\theta / 2) + \displaystyle\frac{Q^2}{ 2|{\vec q}|^2}~, \\[2mm] 
V_{LT} = \displaystyle\frac{Q^2}{ \sqrt{2} |{\vec q}|^2} ~  
\sqrt{ \tan^2(\theta/ 2) + \displaystyle\frac{Q^2} { |\vec{q}|^2}}~,&& 
V_{TT} = \displaystyle\frac{Q^2}{  2 |{\vec q}|^2}~. 
\end{array} 
\label{silcoeff} 
\ee 
 
The nuclear response functions $W_{i}^A$ can be expressed in term of
 nuclear dynamics, once a model for the nuclear current 
operators $J_\mu^A(0)$, appearing in Eq. (A8), is assumed. 
Nowadays, there is no rigorous 
quantum field theory to describe, from first priciples, 
a many body hadronic system. Usually, in electromagnetic processes, 
 the nuclear responses are related  to the 
nucleon responses by some models, the 
simplest one being the {\it impulse approximation} (IA). 
 The IA is based on the following assumptions:
\begin{enumerate}
\item  The nuclear current operator is the sum of one--body 
nucleon operators
\be 
J_\nu^A(Q^2) = \sum_{N=1}^{A} J_\nu^N(Q^2), 
\label{currents} 
\ee 
i.e.
the sum of currents for Dirac 
particles   
 treated within an effective quantum field theory, i.e.
with their internal structure  
described by some phenomenological form factors; therefore the effective 
current operators for nucleons are $Q^2$-dependent and so 
is the nuclear  current operator,
where the $Q^2$-dependence in the r.h.s. and l.h.s. of  
Eq. (\ref{currents}) 
can in principle  differ;
\item the final hadronic state 
$|\alpha_X p_X  , \alpha_B \vec P_B E_B^* \rangle$ 
asymptotically consists of two non interacting (i.e. plane wave states) 
systems, i.e. 
    
\be 
|\alpha_X p_X, \alpha_B \vec P_B E_B^* \rangle =
 {\hat{\cal A}} \{ | \alpha_X p_X  \rangle |\alpha_{B}\vec P_B E_B^*  
 \rangle \} ,
\label{finalA} 
\ee 
where $\hat {\cal A}$ is a proper antisymmetrization operator;
\item the inchoerent contributions leading to the emission of $X$,
due to the interaction of $\gamma^*$ with $B$, are disregarded.
\end{enumerate}

It is straightforward to show that if one adhers to the above assumptions, inserts in (\ref{hadrontz})
a complete set of plane wave nucleon states, and
assumes  the conservation of linear momentum   by using 
traslationally invariant nuclear wave functions, i.e.

\begin{eqnarray}
\langle \alpha_A \vec P_A | \{|\alpha_N\vec p_N  \rangle | \alpha_B \vec P_B 
E_B^* \rangle \}  & = & 
\delta(\vec P_A - \vec p_N - \vec P_B ) 
\delta_{\alpha_A, \alpha_N + \alpha_B}
\nonumber
\\ 
& &\langle \alpha_A\vec P_A  |\alpha_N \vec p_N , \alpha_B\vec P_A-\vec p_N 
E_B^*  \rangle ~,
\label{overlap}
\end{eqnarray}
then in the lab system, the contribution from protons (${t_N = 1/2}$) or neutrons
(${t_N = -1/2}$) 
to the hadronic tensor becomes ($\vec p_N = -{\vec P_B}$):
\begin{eqnarray}
W_{\mu\nu}^{A, s.i., t_N}(\nu, Q^2, \vec p_N)& = &  
{ 1 \over 4 \pi} \frac{M}{E_{\vec p_N}} {1 \over 2}
\overline{\sum_{s_N}}
\sum_{\alpha_X}  
\int d \vec p_X   
\langle {\alpha_N\vec p_N}  | J_\mu^N(0) |  
\alpha_X p_X \rangle \langle \alpha_X p_X | J_\nu^N(0) | 
{\alpha_N\vec p_N}  \rangle
\nonumber
\\
& & 
\delta (M_A + \nu - p_n^0 -p_X^0) ~
\delta (\vec q + \vec p_{N} - \vec p_X) ~ n_{E_B^*}^{t_N}(|\vec p_N|)
\label{wmunuatn}
\end{eqnarray}
where 
\be
n_{E_B^*}^{t_N}(|\vec p_N|) = A {\overline{\sum_{\alpha_{A}}}}
\sum_{\alpha_{B}} 
\left | \langle \alpha_A {\vec P_A}=0 \,|\alpha_N \vec p_N;\alpha_{B} - \vec p_N
E_{B}^*  \rangle \right |^2 \delta_{\alpha_A, \alpha_N + \alpha_B}
\label{npn}
\ee
represents the nucleon momentum distribution  ( assumed to be independent of $s_N$),
corresponding to the intrinsic
 excitation energy $E_{B}^*$ of $B$. Introducing the nucleon spectral function 
\begin{eqnarray}
P_N^{t_N}( |\vec p_N|, E) =
A \overline{\sum_{\alpha_{A}}}
\sum_{\alpha_{B}} \sum_{f}
\left | \langle \alpha_A {\vec P_A}=0\,|\alpha_N \vec p_N; \alpha_{B}- \vec p_N
E_B^f  \rangle \right |^2
\nonumber
\\
\delta(E - (E_B^f - E_A^0)) ~
\delta_{\alpha_A, \alpha_N + \alpha_B} ~, 
\label{ptn}  
\end{eqnarray} 
we can write
\be
n_{E_B^*}^{t_N}(|\vec p_N|)= \int dE P_N^{t_N}( |\vec p_N|,E)\delta_{E_B^f,
E_B^0+E_B^*}
\label{enne}
\ee
where we have considered only the discrete excited states of $B$.
Due to our ignorance of the nucleon current matrix elements in the nuclear tensor (\ref{wmunuatn}), a common practice is to express the latter  in term of the nucleon tensor (Eq. (\ref{nuctz}), $A=N$); however, whereas the three-momentum conservation 
is the same in (\ref{nuctz}) and (\ref{wmunuatn}), 
the energy conservation is not, 
%Eqs. (\ref{wmunuatn})
%and (\ref{nuctz})
being, respectively, $\nu + M_A =
\sqrt { {(M_B+E_B^*)}^2 + \vec {p_N}^2 } - 
\sqrt{ \vec {p_X}^2 + {M_X}^2 }$ in (\ref{wmunuatn}),
and $\nu + E_{\vec p_N} = E_{{\vec p_N} + {\vec q}} $ in (\ref{nuctz}); 
as a result, the nuclear tensor  cannot be directly related to 
the nucleon one, unless some additional, {\it ad hoc} 
assumptions are made. To this end,  
two main prescriptions have been proposed :

\begin{enumerate}

\item 
the hit nucleon is considered to be on-shell, i.e. with a four momentum equal to the one of a free nucleon $p_N ^ {{on}} = ( \sqrt{ 
\vec {p_N}^2   + {M}^2 }, \vec p_N)$ and in (\ref{wmunuatn})
the replacement 
$\nu  \longrightarrow \bar\nu= \nu + M_A -
\sqrt { {(M_B + E_B^*)}^2 + {\vec p_N}^2 } - 
\sqrt{ \vec {p_N}^2 + {M}^2 }$ is done, so that $\delta (M_A + \nu - P_B^0 -p_X^0)  \longrightarrow \delta ( \sqrt{ \vec {p_N}^2 + {M}^2 }+\bar\nu - p_X^0)$ ; 
by this way, the electromagnetic vertex of the nuclear tensor (\ref{wmunuatn})
corresponds to that of a free 
nucleon, 
evaluated at the same ${\vec q}$, but at the transferred 
energy $\bar\nu$ instead of $\nu$ \cite{heller,sauer,forest}, which  
 means that the nucleon hadronic tensor (\ref{nuctz}) has to be evaluated for
$p_N = p_N^{on}$ and $Q_N^2=\bar Q^2 = \vec {q} ^2 - \bar \nu^2 \neq Q^2$.

\item  
The hit nucleon is considered off-shell, with four-momentum
$ p_N^{off} = ( p_N^0, \vec p_N)$ with $p_N^0 = M_A -\sqrt{(M_B + E_B^*)^2 + \vec {p_N}^2}$
and $\delta (M_A + \nu - P_B^0 -p_X^0)  \longrightarrow \delta (p_N^0+ \nu - p_X^0)$, which means that the nucleon hadronic tensor (\ref{nuctz})
 has to be evaluated for
$p_N = p_N^{off}$ and $Q_N^2=\vec q ^2 - \nu^2 = Q^2$.

\end{enumerate}
 In both cases the nuclear tensor  (\ref{wmunuatn}) can be
expressed through the nucleonic tensor ((\ref{nuctz}) with $(A=N)$) obtaining:
\be
&&
W_{\mu\nu}^{A,s.i.t_N}(\nu, Q^2,\vec p_N) =
W_{\mu\nu}^N(p_N \cdot q, Q_N^2,p_N^2)
\frac{M}{E_{\vec p_N}} n_{E_B^*}^{t_N}(|\vec p_N|)
\label{ross1}
\ee

As discussed in details in Ref. \cite{sauer}, both choices imply the presence of many-body currents, due to the dependence of $p_n^{off}$ and $\bar\nu$ upon
the four-momentum of the nucleus. There it has also been stressed that the instant-form dynamics (used in the present paper)
does not mandate one or the other choice. A comparison of both procedures will be presented elsewhere (\cite{bssemi}); 
in the present paper choice $2$ has been adopted, and the $Q^2$ dependence of the  
hadronic tensor for an off mass shell nucleon is assumed to be  the same 
as for the  free one, i.e. 
 \be 
    W_{\mu \nu}^{t_N}(p_N \cdot q, Q_N^2) = - W_1^{N}(p_N \cdot q, Q_N^2) \left [ g_{\mu \nu} + 
    {q_{\mu} q_{\nu} \over Q^2} \right ] + {W_2^{N} (p_N \cdot q, Q_N^2) \over M^2} 
    \tilde{p}_{\mu} ~ \tilde{p}_{\nu} .
    \label{incN} 
 \ee 
Inserting (\ref{incN}) into (\ref{ross1}) and 
the latter into (\ref{wll1}), one gets for the nuclear response functions
(\ref{response}):
 \be 
    W_i^{A}(\nu, Q^2,\vec p_N) = {M \over E_{\vec p_N}} n_{E_B^*}(|\vec p_N|)  
  \sum_{\alpha = 1,2} 
     C_i^{\alpha}(\nu, Q^2, p_N) ~ W_{\alpha}^{N}(p_N \cdot q, Q_N^2) 
    \label{3} 
\ee 
with the   
 coefficients $C_i^{\alpha}(\nu, Q^2, p_N)$ straightforwardly  
obtained from Eqs. (\ref{response}) and  (\ref{silcoeff}), viz. 
  \be 
    C_L^1 = - {|\vec{q}|^2 \over Q^2} ~~~~ 
    C_L^2 & = & {|\vec{q}|^4 \over Q^4} \left [ { p_N^0 |\vec{q}| + 
    \nu |\vec P_{B}|  cos(\theta_{\widehat{\vec P_{B} \vec q}} )  
    \over M |\vec{q}|} \right ] ^2 \nonumber \\ 
    C_T^1 = 2 ~~~~~~~~~ 
    C_T^2 & = & \left ( {  |\vec P_{B}| sin(\theta_{\widehat{\vec P_{B} \vec q}}   )  
    \over M} \right ) ^2 \nonumber \\ 
    C_{LT}^1 = 0 ~~~~~~~ 
    C_{LT}^2 & = & {|\vec{q}|^2 \over 
    Q^2} {\sqrt{8} |\vec p_{N}| sin(\theta_{\widehat{\vec P_{B} \vec q}} )  
    \over M} {p_N^0 |\vec{q}| + \nu  |\vec P_{B}|  
    cos(\theta_{\widehat{\vec P_{B} \vec q}} )  
    \over M |\vec{q}|}  
    ~ cos(\phi_{B}) \nonumber \\ 
    C_{TT}^1 = 0 ~~~~~~~ 
    C_{TT}^2 & = & {1 \over 2} \left ( { |\vec P_{B}|  
    sin ( \theta_{ \widehat{\vec P_{B} \vec q} } )  
    \over M} \right )^2 
    cos(2\phi_{B}) 
    \label{6} 
\ee     
where $\phi_{B}$ is the azymutal angle of $\vec P_B$. Changing variables from $\nu$, $Q^2$, and $(p_N \cdot q)$ to $x_{Bj}=Q^2/(2M\nu)$, $Q^2$, and $x_A=Q^2/2(p_N \cdot q)$, introducing the usual structure functions $F_1 = M W_1$
and  $F_2 = {p_1 \cdot q \over M}  W_2$, using
the Callan--Gross relation, $F_2=2xF_1$, and 
placing Eq. (\ref{3}) into  Eq. (\ref{1}), eq. (\ref{crosa-1}) is obtained,
where $p_N \equiv p_1$ and $B=(A-1)$.

The hadronic tensor for the second process we are considering,
 viz. $A(e,e'B_1B_2)X$,
with $B_1=N_1$ and $B_2= (A-2) $,  will  depend upon six
response  functions which are given by proper bilinear combinations
of the Fourier transforms of the transition matrix elements of 
the nuclear current operator (see e.g. \cite{boffi}). By following the procedure
described above and by introducing the two-nucleon spectral function (\ref{spectr2}),
Eq. (\ref{crosa-2}) can be readily obtained.

%\end{document}

%%%%%%%% CAPTIONS %%%%%%%    
 
\newpage 
\begin{figure} %Fig. 1a and Fig 1b   
\caption{ The processes $A(e,e'(A-1))X$ ) $(a)$ and $A(e,e'N_2 (A-2))X$  
$(b)$ within the Impulse Approximation ( the { \it Spectator mechanism}).} 
\label{fig1} 
\end{figure} 
 
\begin{figure} %Fig. 2a and Fig 2b   
\caption{  
    $(a)$ The ratio  
    $R(|{\vec P_{A-1}}|)$  
    =  
    $\frac{\sigma [D(e,e'p)X]}{\sigma [A(e,e' (A-1))X]}$ 
    ( Eq. (32)) for different nuclei A, viz $^3H$, $^4He$ and    
 $^{12}C$, assuming that DIS took place on a 
neutron . The ratio is plotted versus the value of the momentum  
$P_{A-1}$ $\equiv {\left | {\vec P_{A-1}}\right |}={\left | {\vec P_{A'-1}}\right |}$ of the nucleus 
 emitted backward. $(b)$  The same as in $(a)$  but  
on a linear plot for the targets $^{3}H$ and $^{4}He$ (in this and  
in the following Figures   
$\theta_{P_{A-1}} \equiv \theta_{  \widehat{ \vec P_{A-1} \vec q}}$). }   
\label{fig2} 
\end{figure}

\begin{figure} %Fig. 3   
\caption{ The ratio of the kinematical factor  
$K^A( x_{Bj},Q^2,y_A,z_1^{(A)})$,  (Eq. (\ref{ka})), for $^{4}He$,  
to the same quantity for a free nucleon  $K^N( x_{Bj},Q^2,y)$,  
(Eq. \ref{kap}), vs. $Q^2$ for two values of $x_{Bj}$. } 
\label{fig3} 
\end{figure} 
 
\begin{figure} %Fig. 4   
\caption{ The cross section for the process $A(e,e'(A-1))X$,   
(Eq. (\ref{crosa-1})) on different targets with the nucleus $(A-1)$ 
emitted backward with momentum $P_{A-1}$  
$\equiv {\left | {\vec P_{A-1}}\right |}$ .} 
\label{fig4} 
\end{figure} 
 
\begin{figure} %Fig. 5  
\caption{ The ratio  
  $R^A(x_{Bj},x_{Bj}',z_1^{(A)},Q^2)$ ( Eq. (\ref{x1x2})) 
 for $A=2$ and $A=40$ , $x_{Bj}$ = 0.2 and  $x_{Bj}'$ = 0.5,  $Q^2$= 20 ${GeV/c}^2$, 
plotted versus  the momentum of the nucleus $(A-1)$ emitted at $90^o$ 
. The full and   dashed curves were obtained within the binding  
(x-rescaling) and $Q^2$-rescaling models, respectively.} 
\label{fig5} 
\end{figure} 
 
\begin{figure} %Fig. 6 
\caption{The same as in Fig.5 for nuclei $(A-1)$ emitted backward.} 
\label{fig6} 
\end{figure} 
 
\begin{figure} %Fig. 7 
\caption{The ratio  (45), calculated at $Q^2$ = 20 ${GeV/c}^2$,
$|\vec P_{A-2}|$ = $|\vec P_{A'-2}|$ = 50 $MeV/c$ , 
$x_{Bj}$ = $0.4$,  $\theta_{ \widehat{ \vec p_{2} \vec P_{A-2}}}$ =  
$\theta_{  \widehat{ \vec p_{2} \vec P_{A'-2}}}$ = $10^o$, 
for $A= 12, 40$ and $56$ and $A' = 4$, versus 
$|\vec p_{rel}|$ = $|\vec p_2 + \vec P_{A-2}|$ ($|\vec p_2|$ is varied).} 
\label{fig7} 
\end{figure} 
 
\begin{figure} %Fig. 8 
\caption{The ratio $R_2$ (Eq.(49)), for $^{12}C$ versus  
$x_{Bj}$ for fixed values of: $ i)$ the momentum of the recoiling  
$(A-2)$ system $ P_{A-2} \equiv |\vec P_{A-2}|$ = 50 ${MeV/c}$; $ii)$  
the momentum of the recoiling nucleon $N_2$,  $p_2 \equiv |\vec p_2|$ =  
400 $MeV/c$ (a) and 500 $MeV/c$ (b); $iii)$  
the emission angle of nucleon $N_2$ ($\theta_{p_2}  
\equiv \theta_{  \widehat{ \vec p_{2} \vec q}}$ = $90^o$). The full  
lines correspond to the binding (x-rescaling) model and the dotted lines  
to the $Q^2$-rescaling model with explicit off-shell dependence of the  
nucleon structure function.} 
\label{fig8} 
\end{figure} 
 
\begin{figure} %Fig. 9 
\caption{The inclusive  local EMC effect in $^{12}C$. The full curve  
represents the inclusive EMC ratio due to the mean field nucleons in  
$^{12}C$, i.e. Eq. (\ref{r0}) integrated over all space  
($0\leq P_{A-1}\leq \infty$, $0 \leq \theta_{p_{A-1}} \leq \pi$), whereas  
the dashed and dotted lines represent the contribution from  $1p$  
and $1s$-shell nucleons, respectively.} 
\label{fig9} 
\end{figure} 
 
\begin{figure} %Fig. 10 
\caption {The seminclusive EMC ratio ${\sigma (^{12}C})/{\sigma (D)}  
\equiv R_{o} (x_{Bj}, Q^2)$, (Eq. (\ref{r0})) corresponding to nuclei  
emitted backward and forward, in the kinematical ranges shown in the Figure.  
The full curve is the usual inclusive EMC ratio.} 
\label{fig10} 
\end{figure} 
 
\begin{figure} %Fig. 11 
\caption{The backward seminclusive local EMC effect on $^{12}C$ i.e. the  
contribution to the ratio $R_o$ (Eq. (\ref{r0})) of the nuclei $(A-1)$  
emitted backward in the range $160^{o} \leq \theta_{A-1} \leq 180^{o}$,   
$P_{A-1} \leq {2 fm^{-1}}$. The dashed curve represents the usual inclusive  
EMC ratio 
(Eq. (\ref{r0}), integrated over all space).} 
 \label{fig11} 
\end{figure} 
  
\begin{figure} %Fig. 12 
\caption{The seminclusive cross section (Eq. (\ref{ka})) resulting from DIS  
on  $s$-shell (dashed) and $p$-shell (full) nucleons 
of $^{12}C$. The results are plotted versus the emission angle  
$\theta_{P_{A-1}} \equiv \theta_{  \widehat{ \vec P_{A-1} \vec q}}$  
of the recoiling $(A-1)$ nuclei, for fixed value of $x_{Bj} \equiv x$ and  
in correspondence of two values of the momentum $P_{A-1} \equiv  
|\vec P_{A-1}|$ of the recoiling  $(A-1)$ nucleus.} 
\label{fig12} 
\end{figure} 
 
\newpage 
%%%%%%%%% Insertion of figures %%%%%%%%%%%%%% 
 
%%                                   Fig1a 
 
\epsfxsize 8cm  
  \hspace*{-2cm}\epsfbox{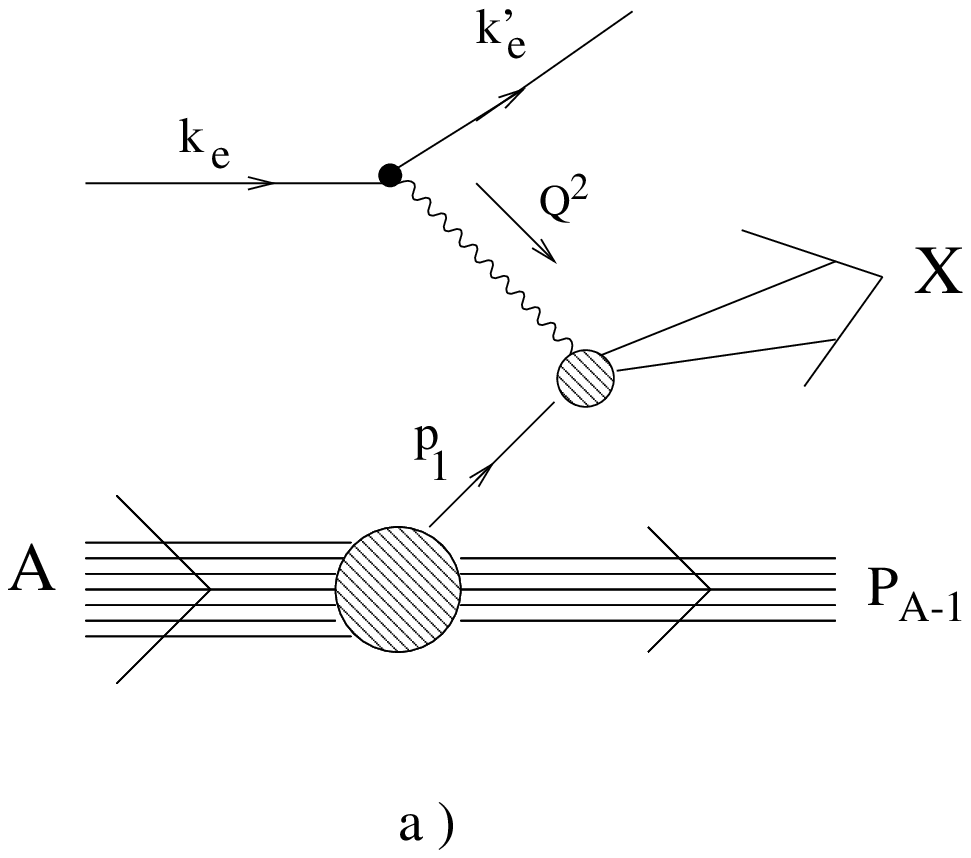} 
 
\vskip -7.5cm 
%-----------------------Fig.1b 
\epsfxsize 8cm      
\hfill\epsfbox{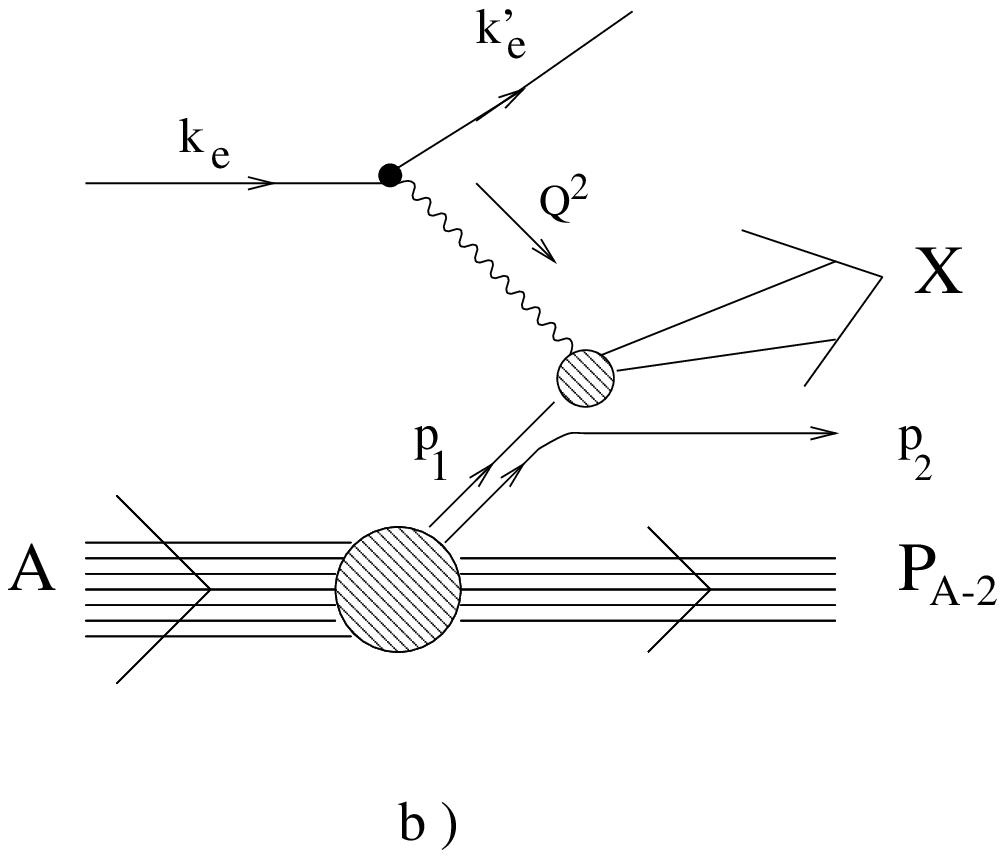} 
 \vskip 2.2cm 
\hspace{3cm}   
\vfill   
 
Fig.~\ref{fig1}.  
C. Ciofi degli Atti.... Seminclusive deep inelastic...

 \newpage 
%%----------------   Fig.2 --------------------- 
  
\epsfxsize 9cm      
  \hspace*{-2cm}\epsfbox{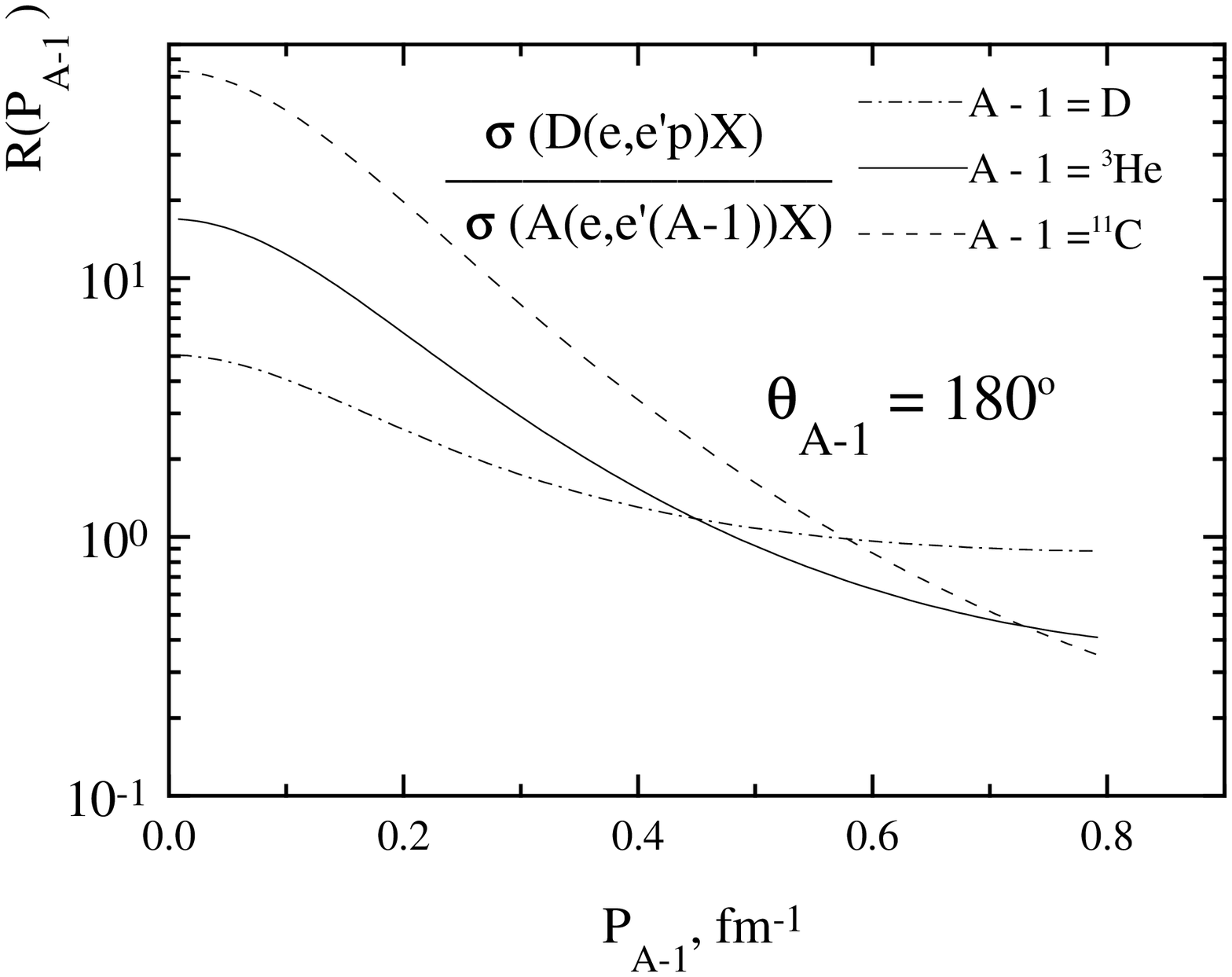} 
 
\vskip -8.cm 
 
   %-----------------------Fig.2b 
\epsfxsize 8cm      
\hfill\epsfbox{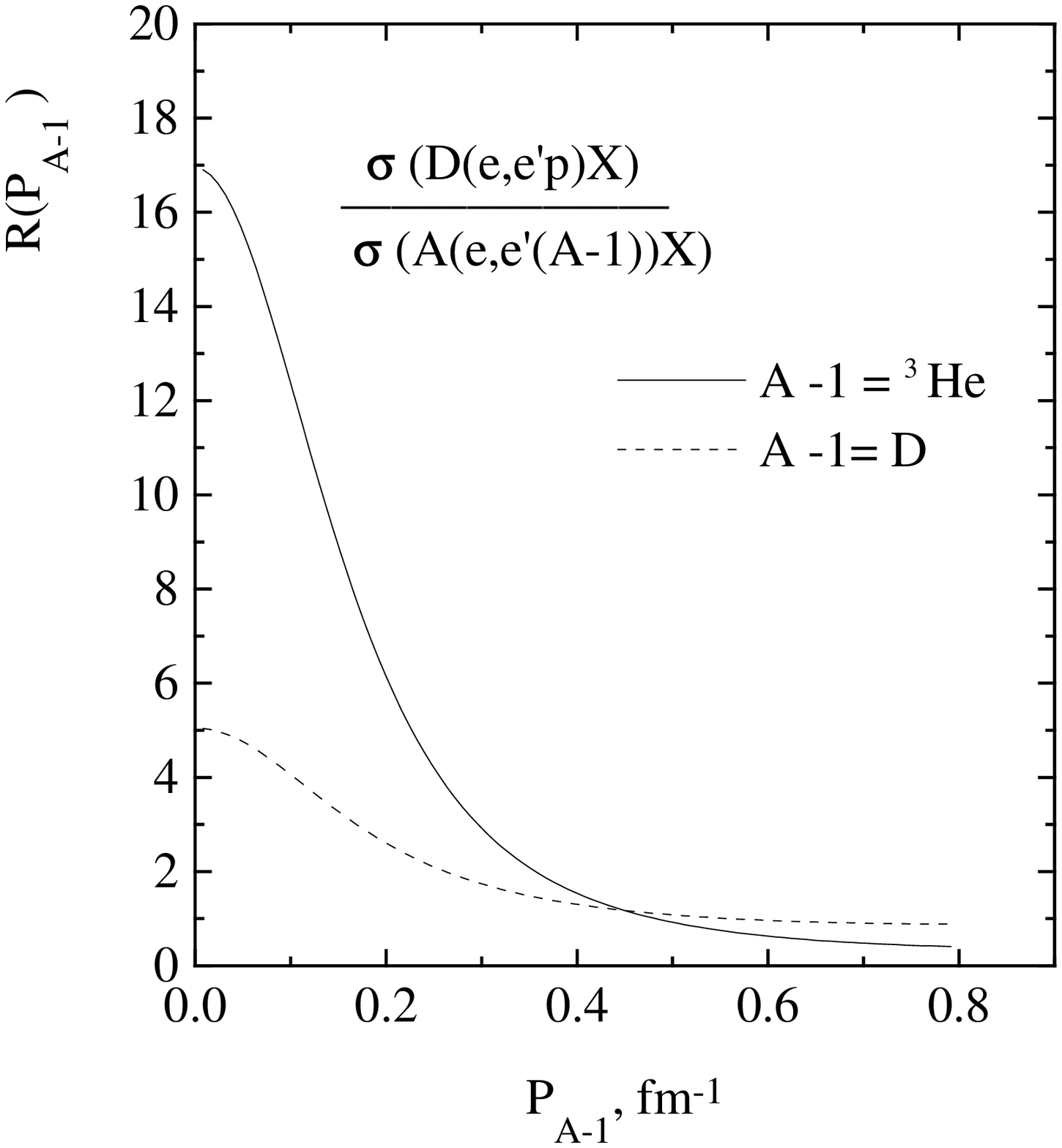} 
 \vskip 2.2cm 
\hspace{3cm} a) \hspace{10cm} b) 
\vfill 
 
Fig.~\ref{fig2}.  
C. Ciofi degli Atti.... Seminclusive deep inelastic... 
 
\newpage 
   %-----------------------Fig.3 
\epsfxsize 12cm      
\centerline{\epsfbox{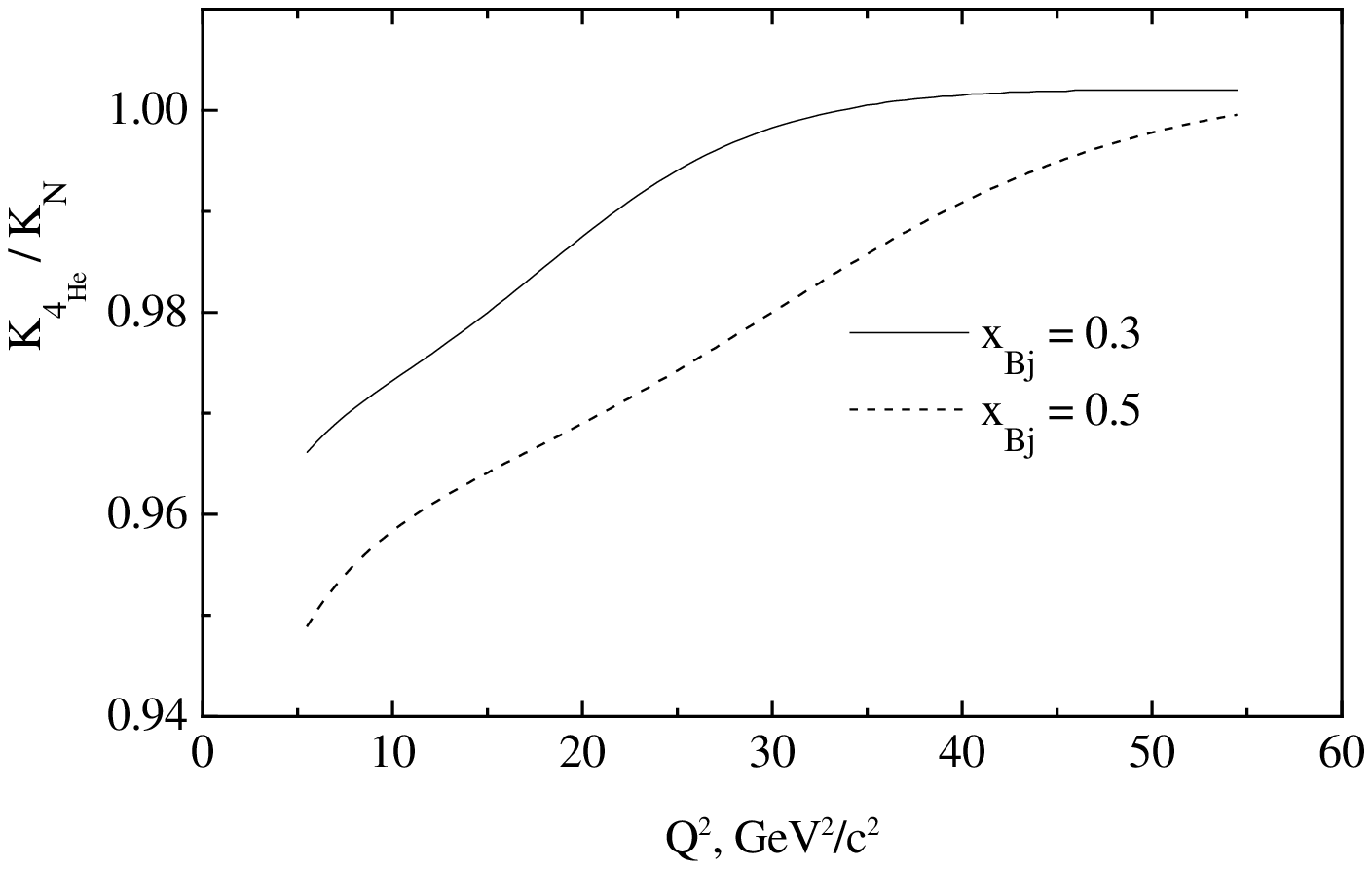}} 
  \vfill 
   
Fig.~\ref{fig3}.  
C. Ciofi degli Atti.... Seminclusive deep inelastic... 
 
\newpage 
   %-----------------------Fig.4 
\epsfxsize 12cm      
\centerline{\epsfbox{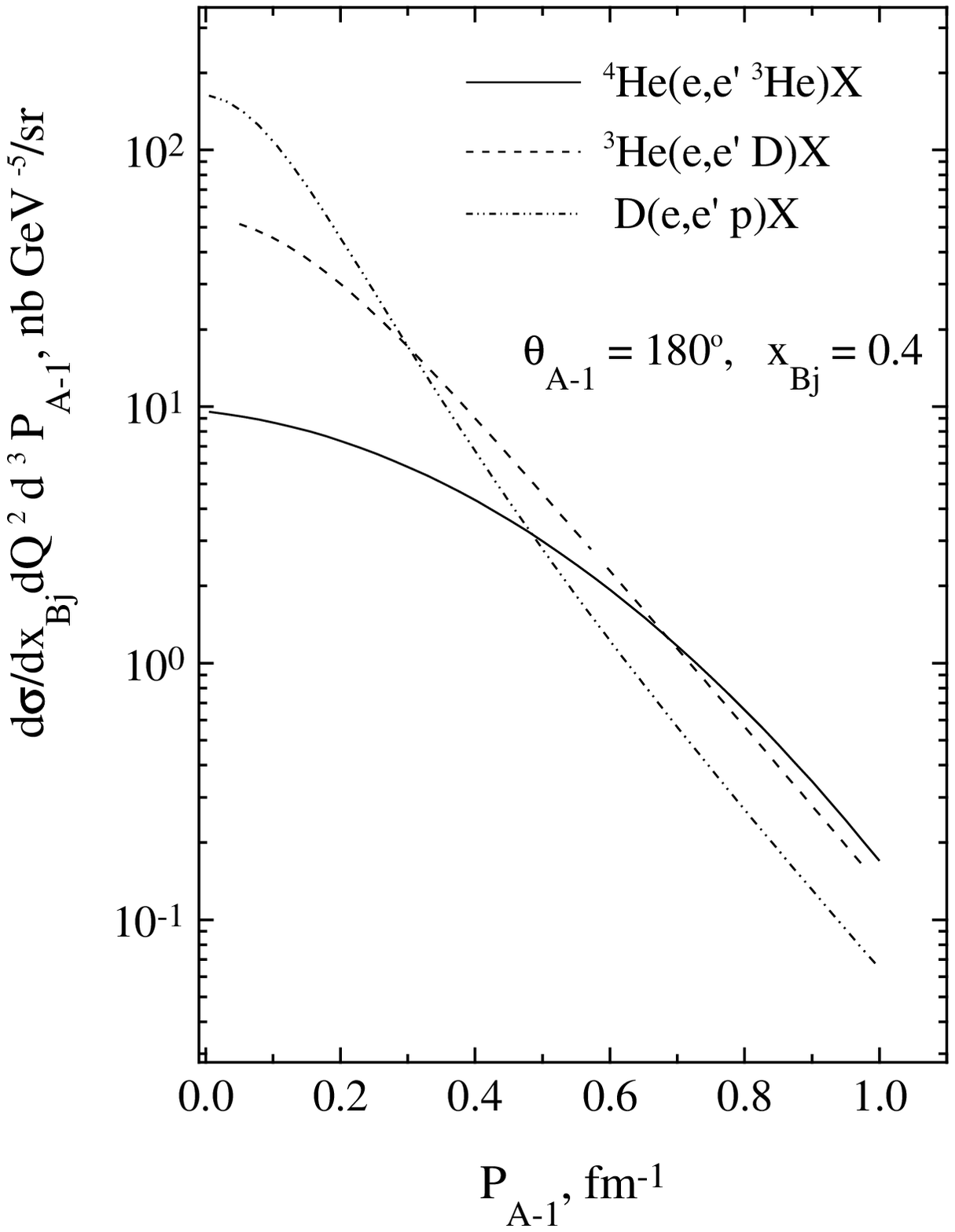}} 
  \vfill 
   
Fig.~\ref{fig4}.  
C. Ciofi degli Atti.... Seminclusive deep inelastic...

\newpage 
   %-----------------------Fig.5 
\epsfxsize 12cm      
\centerline{\epsfbox{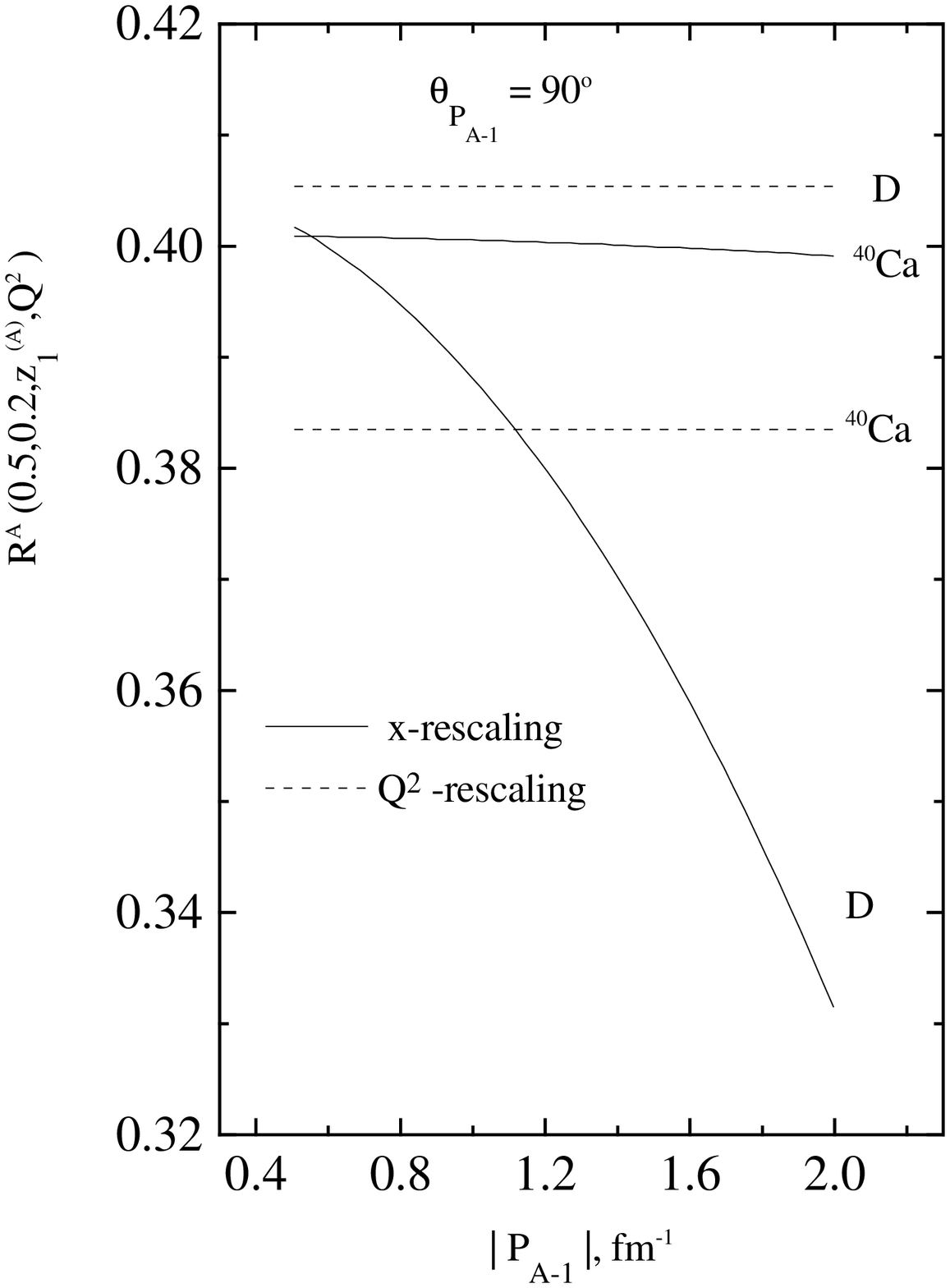}} 
  \vfill 
   
Fig.~\ref{fig5}.  
C. Ciofi degli Atti.... Seminclusive deep inelastic...

\newpage 
   %-----------------------Fig.6 
\epsfxsize 12cm      
\centerline{\epsfbox{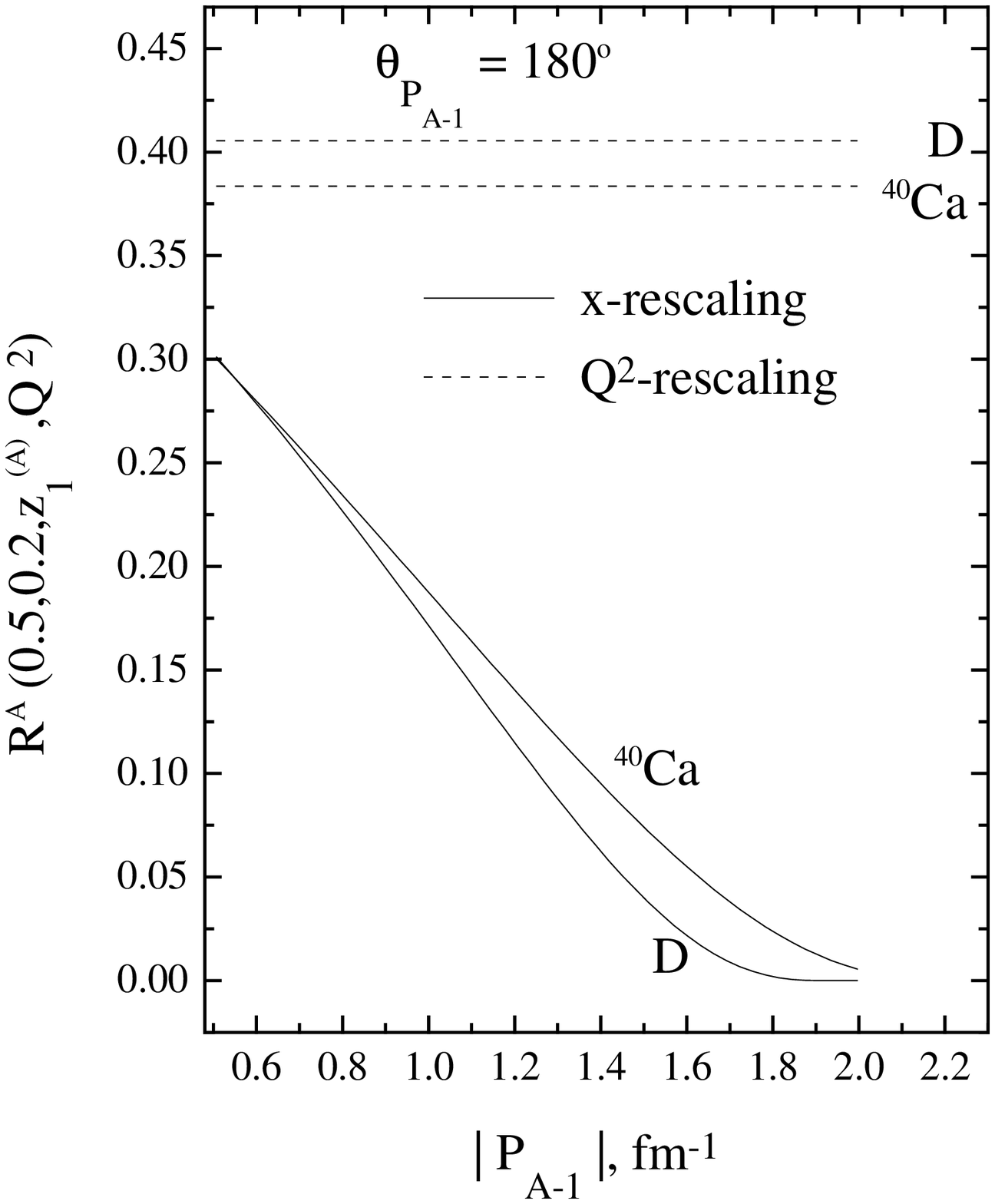}} 
  \vfill 
   
Fig.~\ref{fig6}.  
C. Ciofi degli Atti.... Seminclusive deep inelastic...

\newpage 
   %----------------------Fig.7 
\epsfxsize 12cm      
\centerline{\epsfbox{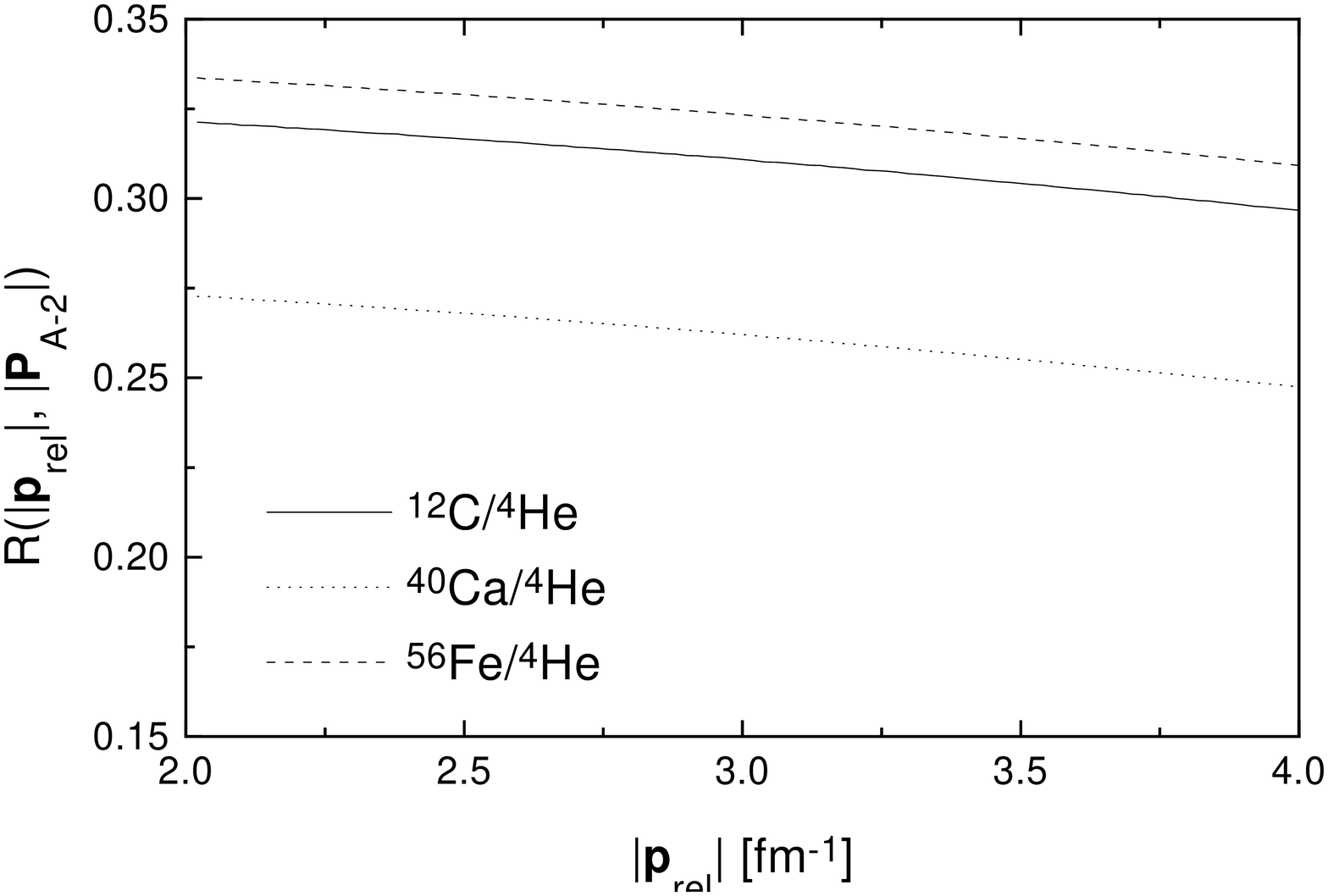}} 
  \vfill 
   
Fig.~\ref{fig7}.  
C. Ciofi degli Atti.... Seminclusive deep inelastic...

\newpage 
   %----------------------Fig.8 
\epsfxsize 12cm      
\centerline{\epsfbox{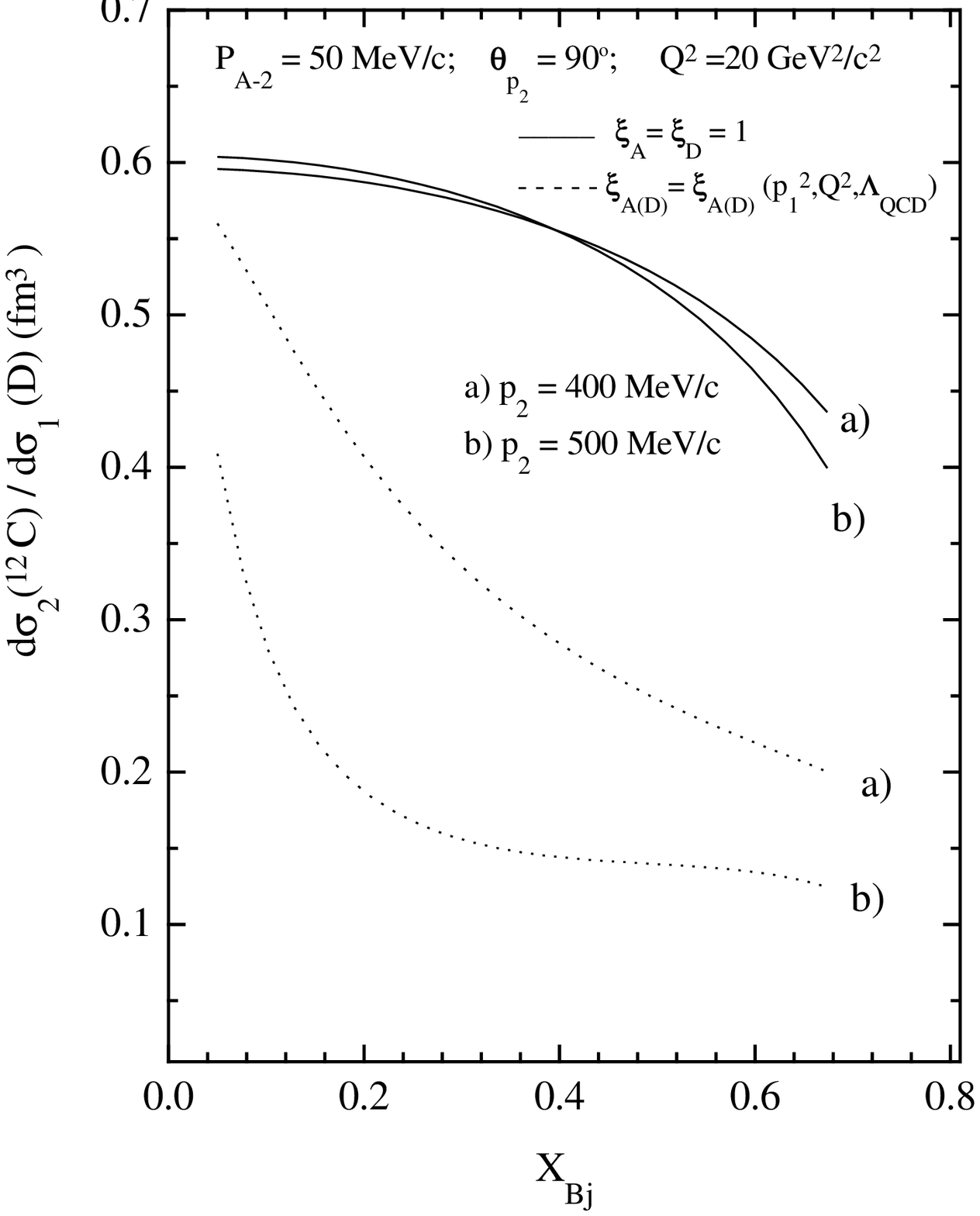}} 
  \vfill 
   
Fig.~\ref{fig8}.  
C. Ciofi degli Atti.... Seminclusive deep inelastic...

\newpage 
   %-----------------------Fig.9 
\epsfxsize 12cm      
\centerline{\epsfbox{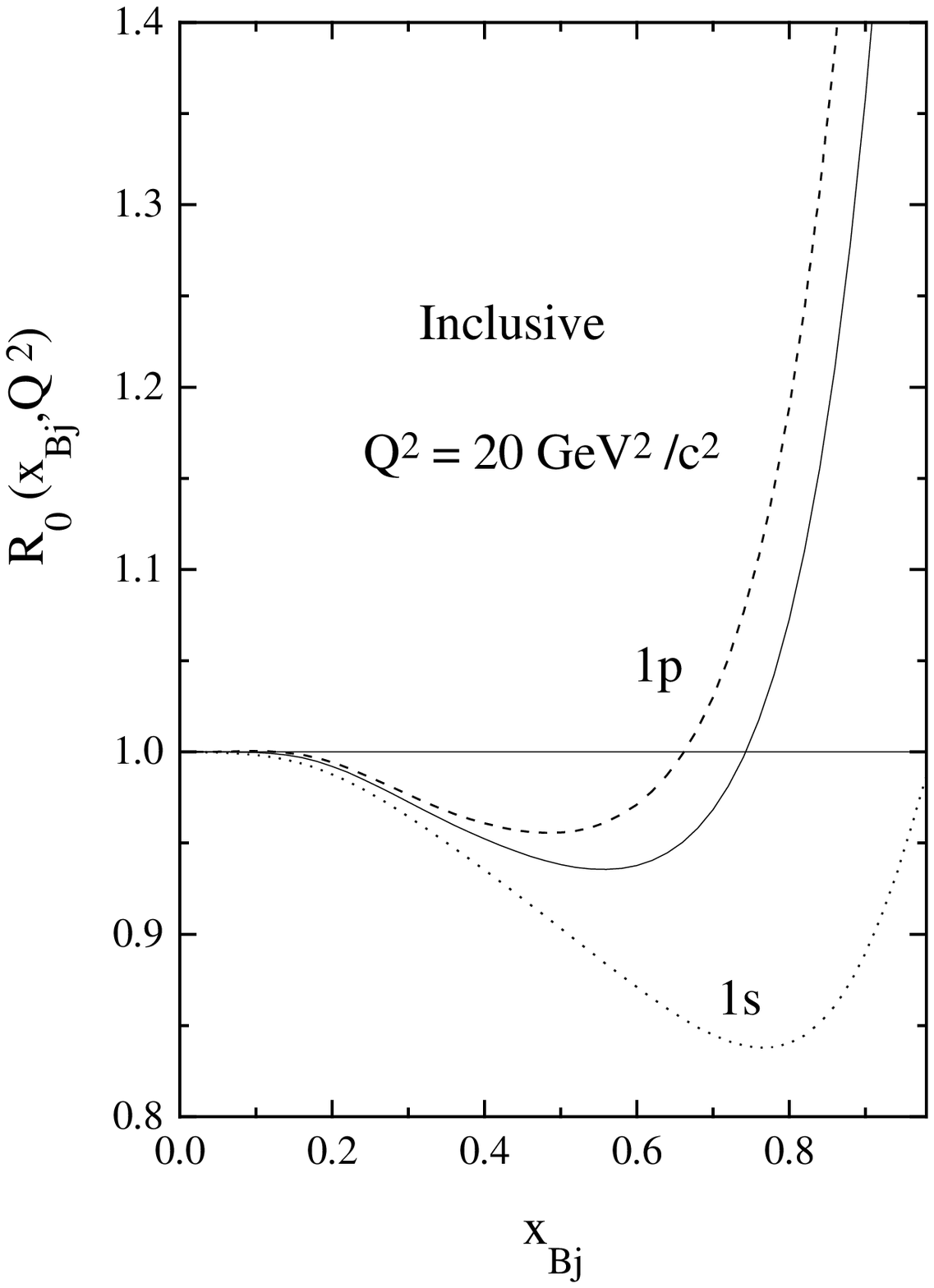}} 
  \vfill 
   
Fig.~\ref{fig9}.  
C. Ciofi degli Atti.... Seminclusive deep inelastic...

\newpage 
   %-----------------------Fig.10 
\epsfxsize 12cm      
\centerline{\epsfbox{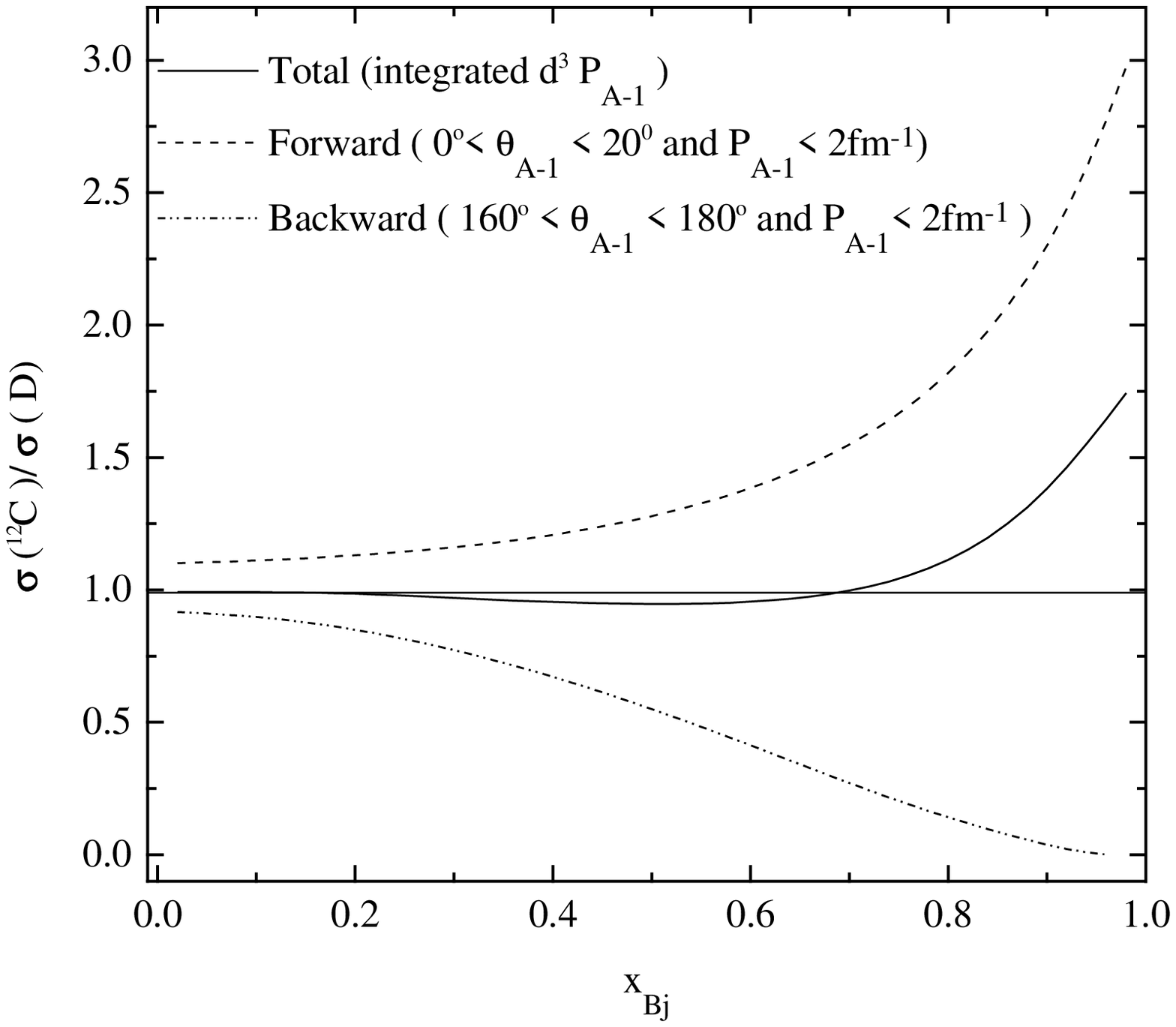}} 
  \vfill 
   
Fig.~\ref{fig10}.  
C. Ciofi degli Atti.... Seminclusive deep inelastic...

\newpage 
   %-----------------------Fig.11 
\epsfxsize 12cm      
\centerline{\epsfbox{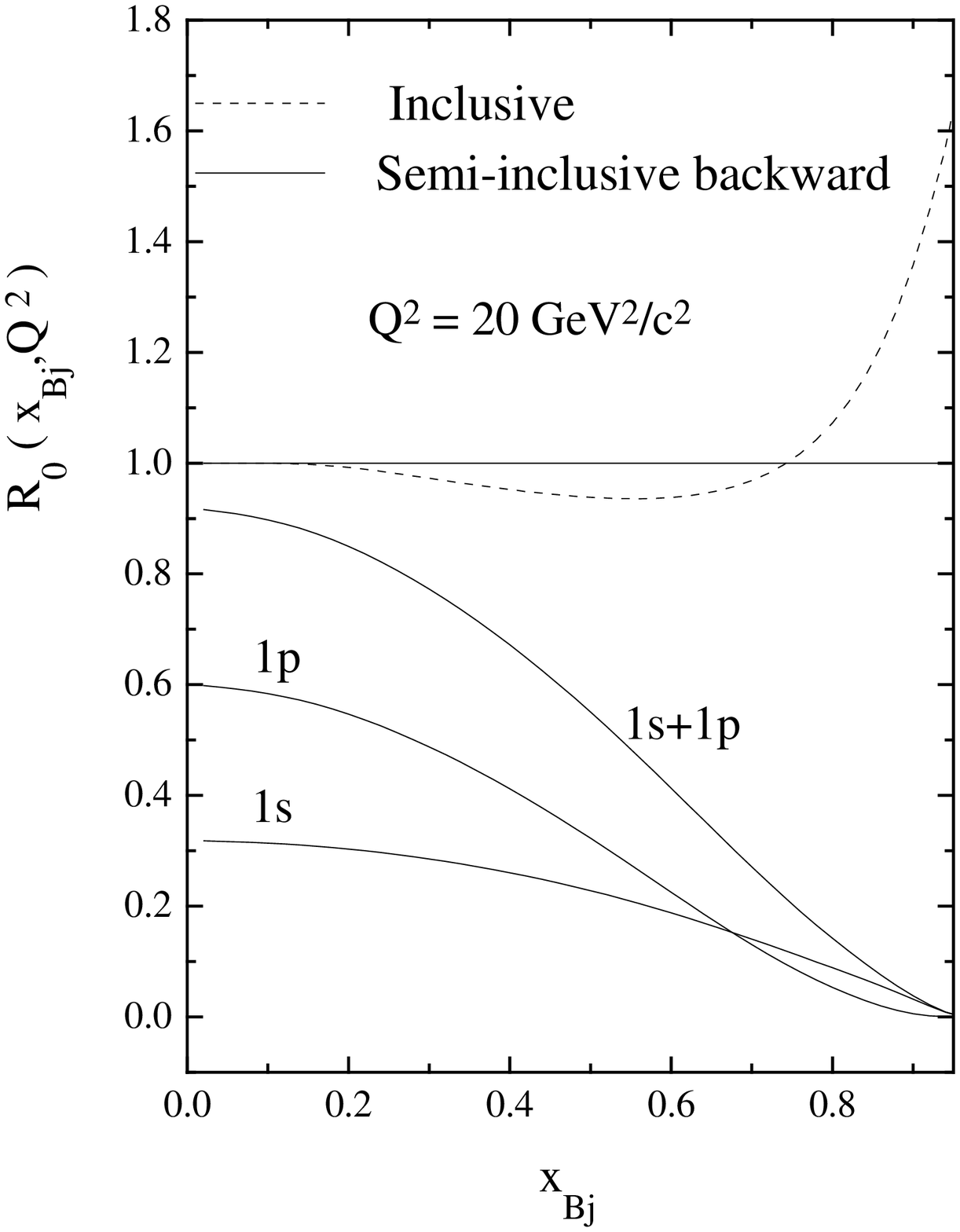}} 
  \vfill 
   
Fig.~\ref{fig11}.  
C. Ciofi degli Atti... Seminclusive deep inelastic...

\newpage 
   %-----------------------Fig.12 
\epsfxsize 12cm      
\centerline{\epsfbox{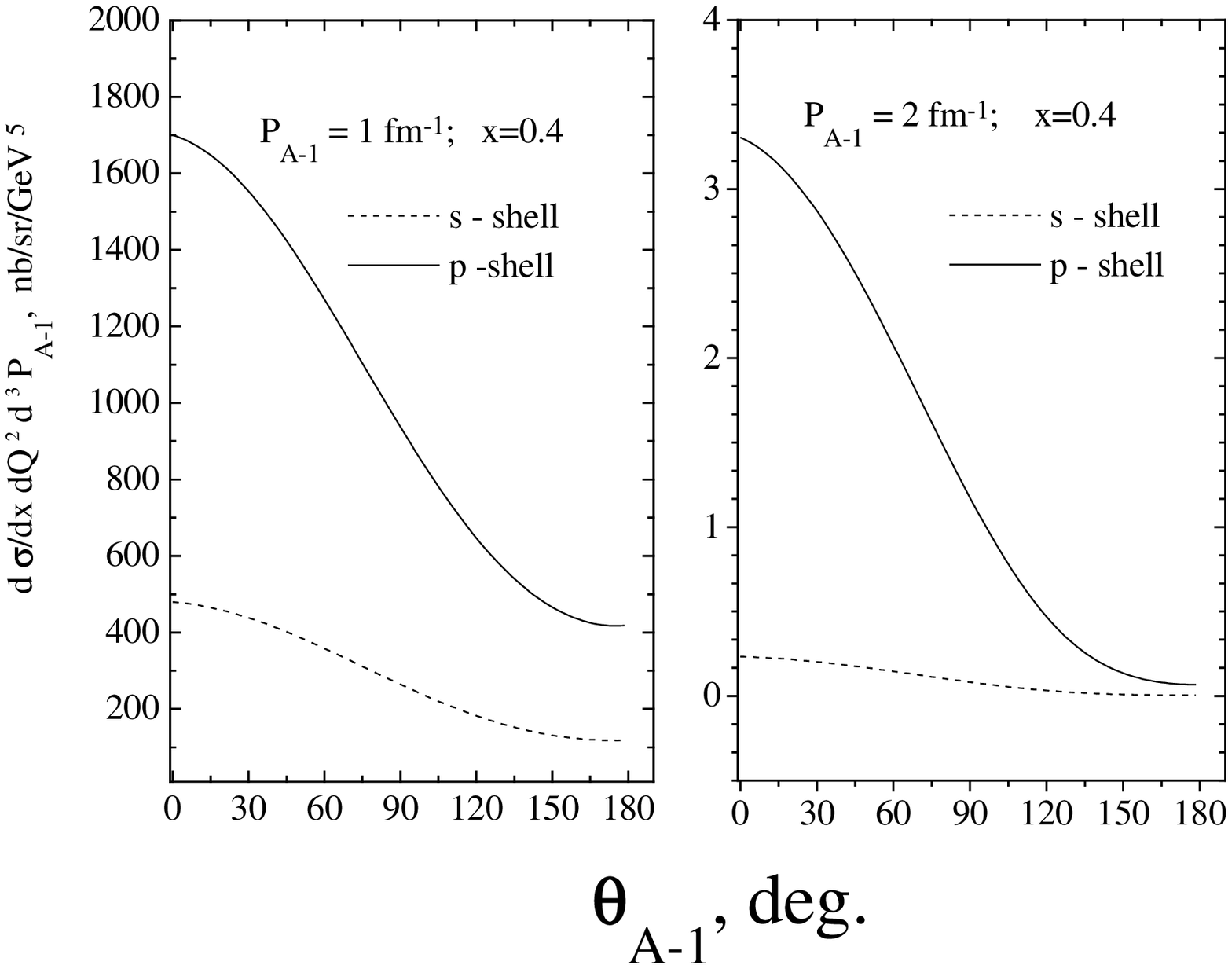}} 
  \vfill 
   
Fig.~\ref{fig12}.  
C. Ciofi degli Atti... Seminclusive deep inelastic...

\end{document}